\documentstyle[prd,aps,eqsecnum]{revtex} 

\newcommand{\be}{\begin{equation}}
\newcommand{\ee}{\end{equation}}
\newcommand{\bea}{\begin{eqnarray}}
\newcommand{\eea}{\end{eqnarray}}

\def\thefootnote{\fnsymbol{footnote}}

\newcommand{\artref}[5]{{\rm #1}, {\rm #2} {\bf #3}, {\rm #4 (#5)}}

\newcommand{\bookref}[3]{{\rm #1}, {\it #2} {\rm (#3)}}

\def\B{B}
\def\Z{{\cal Z}}
\def\W{{\cal W}}

\def\gp{{g^\prime}}
\def\gs{g^2}
\def\gps{\gp^2}
\def\p{\partial}
\def\jmunut{(\p_\mu J_{\nu}^\Z - \p_\nu J_{\mu}^\Z)}

\def\JRmu{{\cal J}_{R\mu}}
\def\lag{{\cal L}}

\def\polemr{\Lambda_\varepsilon(2\mrs)}
\def\lr{\lambda_r}
\def\gr{g_r}
\def\gpr{{g_r^\prime}}
\def\grs{g_r^2}
\def\gprs{\gpr^2}
\def\mrs{m_r^2}

\def\Y{{\cal Y}}

\def\A{{\cal A}}
\def\PT{{\rm PT}}
\def\PL{{\rm PL}}
\def\PTT{\widetilde{\rm PT}}

\def\tU{\widetilde U}
\def\g{g}

\def\tq{q}
\def\tvarD{\widetilde\varD}
\def\hvarD{\widehat\varD}
\def\u{\underline}

\def\varD{{\cal D}}

\def\half{{1\over 2}}

\def\y{\Y^\Z}

\def\V{{\cal V}}     
\def\UoneParam{\omega}  
\def\Bphi{\Phi}
\def\tBphi{\widetilde\Phi}
\def\polem{\Lambda_\varepsilon(2m^2)}
\def\tr{{\rm tr}}
\def\vT{{\widehat T}}
\def\aT{{\widetilde T}}
\def\source{\chi}  
\def\order{{\cal O}}

\def\G{{\cal G}}  
\def\ddp{{d^d p \over (2\pi)^d}} 
\def\ddq{{d^d q \over (2\pi)^d}} 
\def\ers{{e_{\mathrm{res}}^2}}

\def\d{{\rm d}}   
\def\intdx{\int\d^d x\/}
\def\intdy{\int\d^d y\/}
\def\Tr{{\rm Tr\/}}

\def\Uoneem{U(1)_{\mathrm{em}}}
\def\phase{{\varphi}}

\def\SM{{\mathrm SM}}

\def\Allfields{{\cal F}}
\def\cAllfields{{{\cal F}^{cl}}}

\def\gaugeparam{{\omega}}
\def\tildeD{{\widetilde D}}
\def\Trafo{{\cal T}}

\def\bvarD{\bar{\varD}}

\def\DeltaEoM{{T}}

\def\w{w} 
\def\b{b}

\def\tY{{\widetilde \Y}}

\def\MHps{M_{H,\mathrm{pole}}^2}  
\def\MWps{M_{W,\mathrm{pole}}^2}
\def\MZps{M_{Z,\mathrm{pole}}^2}

\def\bD{{\bar D}}
\def\bsigma{{\bar \sigma}}
\def\bGamma{{\bar \Gamma}}
\def\bP{{\bar P}}
\def\bd{{\bar d}}
\def\bvarD{{\bar \varD}}

\def\Dloc{{D_{\mathrm{loc}}}}
\def\Dlocinvers{{D^{-1}_{\mathrm{loc}}}}
\def\dDnl{{\delta D}}
\def\Dfull{{\tildeD + \bP \bP^T}}  
\def\uDfull{{\underline{\tildeD} + \underline{P P^T} + \delta_P}}

\def\uW{{{\underline \W}}}
\def\uuW{{\underline{\underline \W}}}

\begin{document}


\draft

\title{Gauge-invariant Green's functions for the bosonic sector of the
standard model} 

\author{Andreas Nyf\/feler\footnote{Present address: Centre de Physique
    Th\'{e}orique, CNRS-Luminy, Case 907, F-13288 Marseille Cedex 9,
    France; Email address: nyf\/feler@cpt.univ-mrs.fr}}
\address{DESY, Platanenallee 6, D-15738 Zeuthen, Germany} 

\author{Andreas Schenk} 
\address{Hambacher Stra\ss e 14, D-64625 Bensheim-Gronau, Germany} 
    
\date{19 June 2000} 

\maketitle

\begin{abstract}
There are many applications in gauge theories where the usually
employed framework involving gauge-dependent Green's functions leads
to considerable problems. In order to overcome the difficulties
invariably tied to gauge dependence, we present a manifestly
gauge-invariant approach. We propose a generating functional of
appropriately chosen gauge-invariant Green's functions for the bosonic
sector of the standard model.  Since the corresponding external
sources emit one-particle states, these functions yield the same
$S$-matrix elements as those obtained in the usual framework.  We
evaluate the generating functional for the bosonic sector of the
standard model up to the one-loop level and carry out its
renormalization in the on-shell scheme. Explicit results for some
two-point functions are given. Gauge invariance is manifest at any
step of our calculation. 
\end{abstract}

\pacs{PACS number(s): 11.15.Bt, 11.15.Ex, 12.15.-y} 

\renewcommand{\thefootnote}{\arabic{footnote}}
\setcounter{footnote}{0}



\section{Introduction}
\label{sec:intro}

The concept of gauge symmetry has played an important role in the
development of quantum field theory and particle physics.
Nevertheless, most practical calculations in perturbation theory are
done by fixing a gauge at the beginning and using gauge-dependent
Green's functions~\cite{Faddeev_Popov,BRST}. Gauge dependence
manifests itself only in the off-shell behavior of these Green's
functions.  Their pole positions and residues, i.e., particle masses,
decay widths, and $S$-matrix elements, are gauge-independent. As long
as one is only dealing with physical, and therefore, gauge-invariant
quantities there seems to be no reason why the gauge symmetry should
be manifest throughout the whole calculation.

There are, however, situations where one is interested to gain
information from off-shell quantities or where one is forced to deal
with them. For instance, problems with the gauge-dependent approach
arise when one is dealing with finite width effects of unstable
particles, which is relevant for $W$-boson production at the CERN $e^+
e^-$ collider LEP2 or at future colliders~\cite{W-pair-production}.
Another example are the oblique parameters $S,T,U$~\cite{STU} which
parametrize effects of new physics on the vacuum polarization of the
electroweak gauge bosons and which are defined through gauge-dependent
self-energies.  Off-shell information~\cite{EW_phasetransition} is
also used to investigate the electroweak phase transition where one is
studying the effective potential at finite temperature.  In the
context of effective field theories one encounters
gauge-dependencies~\cite{ChPT_em_gaugedep}, if one includes
electromagnetic effects~\cite{ChPT_em} in chiral perturbation
theory~\cite{ChPT}. Another example where a gauge-dependent framework
causes considerable complications is the matching of a full and an
effective theory. As pointed out in Refs.~\cite{E_M,Abelian_Higgs},
there are some subtleties involved concerning gauge invariance, if the
matching is performed at the level of gauge-dependent Green's
functions.

Several attempts have been made in order to solve these problems with
gauge-dependencies in different applications. For instance, the
fermion loop scheme~\cite{fermionloopscheme} was developed to treat
unstable particles in $W$-pair production. The
pinch-technique~\cite{pinch,pinch_recent} was used to define
quantities $S,T,U$ which are independent of the
gauge-parameter~\cite{DKS}.  Another approach to improve the
properties of Green's functions with respect to gauge transformations
is the background field method~\cite{BGFM} which was applied to the
standard model in Ref.~\cite{BGFM_SM}.

The techniques employed differ in the degree with which the symmetry
properties are manifest.  All these approaches work within the usually
employed gauge-fixed framework and try to improve the properties of
Green's functions with respect to the gauge symmetry. Instead, we
proposed a manifestly gauge-invariant functional approach in
Ref.~\cite{Abelian_Higgs} which is better suited for the applications
we have in mind. It deals from the beginning only with Green's
functions of gauge-invariant operators. We first applied our method to
an effective field theory analysis of the Abelian Higgs model.  Later
we showed how one can treat charged particles with this new method in
a manifestly gauge-invariant way~\cite{QED_gaugeinv}. This was done by
a detailed comparison of QED in our approach with the conventional,
gauge-dependent method. The extension to the electroweak standard
model was briefly sketched in that reference as well.

In the present paper we discuss the application of our method to the
non-Abelian case in full detail. Generalizing the functional methods
developed in Refs.~\cite{LSM,Abelian_Higgs,QED_gaugeinv} we
construct a generating functional for appropriately chosen
gauge-invariant Green's functions for the bosonic sector of the
standard model at the one-loop level.  This is done by coupling
external sources to gauge-invariant operators in such a way that the
sources emit one-particle states of the Higgs boson, the $W$- and
$Z$-boson and the photon. Due to the manifestly preserved gauge
symmetry in our approach, the generating functional and the
corresponding Green's functions automatically exhibit useful physical
properties. In the usual approach to gauge theories these properties
have to be imposed by specific renormalization conditions or by
employing Ward identities.  Finally, we show how one can extract
physical quantities like masses, coupling constants and $S$-matrix
elements from these gauge-invariant Green's functions. Since the
external sources emit one-particle states, the gauge-invariant Green's
functions lead to the same $S$-matrix elements as those obtained in
the usual framework. In a first step we have not included fermions in
our analysis. In principle, the treatment of fermions is
straightforward in our approach. The corresponding source terms have
been written down already in Ref.~\cite{QED_gaugeinv}.

Because we couple sources only to gauge-invariant operators it is
possible to calculate the generating functional and the corresponding
Green's functions without fixing a gauge in the path
integral~\cite{Abelian_Higgs,QED_gaugeinv}. At tree level we can solve
the equations of motion for the physical degrees of freedom and
define their propagators. The manifestly gauge-invariant method
generalizes to the one-loop level where no Faddeev-Popov ghost fields
appear. We note that the propagators which enter loop diagrams are not
identical to the ones in unitary gauge in the usual framework.
Therefore, the Green's functions\footnote{Note that we distinguish
between the Green's functions, like two-point or $n$-point functions,
of gauge-invariant operators which are obtained from the generating
functional and the propagators which appear in the solutions of the
equations of motion at tree level or within loop integrals.} of our
gauge-invariant operators have a decent high-energy behavior and the
renormalizability of the theory is clearly visible.  This is due to
the fact that the Goldstone boson modes are still present in the
loops. We will discuss the renormalization of the theory in the
presence of external sources in detail, using dimensional
regularization and employing heat-kernel techniques.

In the present work we concentrate on the development of a
gauge-invariant functional approach to the symmetry breaking sector of
the standard model in the spontaneously broken phase, thereby laying
the theoretical foundations. A first application of our
gauge-invariant method can be found in Ref.~\cite{EWChPT_reanalyzed}
where we have performed a detailed analysis of the electroweak chiral
Lagrangian~\cite{EW_chiral_Lag}, which describes the low-energy
structure of a strongly interacting electroweak symmetry breaking
sector. In Ref.~\cite{EWChPT_reanalyzed} we have studied two issues
related to gauge invariance where the usual approach with
gauge-dependent Green's functions leads to considerable problems. In
particular, we have determined the number of independent parameters in
the effective Lagrangian by making use of the equations of motion to
remove redundant terms. Furthermore, we have evaluated the effective
Lagrangian for the standard model with a heavy Higgs boson. The
calculation was performed by matching gauge-invariant Green's
functions in the full and the effective theory at low energies.

Finally, we would like to stress that the construction of the
generating functional is done in such a way that the gauge symmetry is
manifestly preserved at any stage. In this respect our method differs
from the treatment of charged particles as proposed in
Refs.~\cite{Steinmann,Horan_Lavelle_McMullan}, although the starting
point for the choice of gauge-invariant fields is very similar. We
note that there are other attempts in the literature to define
gauge-invariant and gauge-independent Green's functions in field
theories, for instance the Vilkovisky-DeWitt effective
action~\cite{VDEA}.  Another method which naturally deals with
gauge-invariant objects is of course lattice gauge
theory~\cite{lattice}.

This paper is organized as follows: In the next section we discuss our
choice of gauge-invariant operators and the corresponding external
sources which emit one-particle states of the bosons. Then we define
the generating functional for the Green's functions. In
Sec.~\ref{sec:tree_level} we evaluate the generating functional at
tree level. In particular, we discuss the solutions of the equations
of motion for the physical degrees of freedom. In
Sec.~\ref{sec:one_loop} we calculate the generating functional at
the one-loop level in such a way that the gauge symmetry is manifestly
preserved throughout.  The result encodes all one loop effects of the
theory.  In Sec.~\ref{sec:renorm} we discuss the renormalization of
the model and determine the renormalization prescriptions. In
Sec.~\ref{sec:greensfunctions} we calculate the two-point functions
for the gauge-invariant operators and present some properties of these
Green's functions which follow from the gauge symmetry. Furthermore we
extract the electric charge and the masses of the bosons from the
relevant two-point functions. Finally, we summarize our results in
Sec.~\ref{sec:summary}. Some technical details and lengthy
expressions which are needed for the calculation can be found in
several Appendices.


\section{The Lagrangian and the gauge-invariant  
gen\-er\-at\-ing functional} 
\label{sec:lagrangian}

The Lagrangian of the standard model without fermions is of the form 
\be
\label{Lag_SM_before}
\lag  = 
{1\over2} D_\mu\Bphi^\dagger D_\mu\Bphi 
-{1\over 2} m^2 \Bphi^\dagger\Bphi 
+{\lambda\over4} (\Bphi^\dagger\Bphi)^2 
+ {1\over 4\gs}  W_{\mu\nu}^a W_{\mu\nu}^a 
+ {1\over 4\gps} \B_{\mu\nu} \B_{\mu\nu}  \, , 
\ee
where $\Bphi = \left( \begin{array}{c} \phi^1 \\ \phi^2 \end{array}
\right)$ denotes the Higgs boson doublet which is coupled to the
$SU(2)_L$ gauge fields $W_\mu^a \, (a=1,2,3)$ and the $U(1)_Y$ gauge
field $B_\mu$ through the covariant derivative 
\be
\label{cov_deriv} 
D_\mu \Bphi = \left( \p_\mu - i {\tau^a\over 2} W_\mu^a 
                          - i {1\over 2} B_\mu \right) \Bphi \, .
\ee
Note that we have absorbed the coupling constants $g$ and $\gp$ into
the gauge fields $W_\mu^a$ and $B_\mu$, respectively. The field
strengths are given by 
\bea
W_{\mu\nu}^a  & = & \p_\mu W_\nu^a - \p_\nu W_\mu^a
                        + \varepsilon^{abc} W_\mu^b W_\nu^c \, , \\
B_{\mu\nu}    & = & \p_\mu B_\nu - \p_\nu B_\mu \, . 
\eea
The Higgs field $\Bphi$ and the gauge fields $W_\mu^a, B_\mu$
transform under $SU(2)_L$ gauge transformations in the following way: 
\bea 
\Bphi & \to & \V \Bphi \, , \quad \V \in SU(2) \, , \nonumber \\ 
W_\mu & \to & \V W_\mu \V^\dagger - i (\partial_\mu \V) \V^\dagger  \, ,
\quad W_\mu \equiv W_\mu^a {\tau^a\over 2} \, , 
\label{SUtwo_trafo}
\eea
and under $U(1)_Y$ gauge transformations as follows: 
\bea 
\Bphi   & \to & e^{- i \UoneParam / 2} \ \Bphi \, ,  \nonumber \\
B_\mu & \to & B_\mu - \partial_\mu \UoneParam \ . 
\label{Uone_trafo}
\eea
For computational convenience we are working in Euclidean space-time. 

For $m^2 > 0$ the classical potential has its minimum at a nonzero
value $\Bphi^\dagger \Bphi = m^2 / \lambda$ and the $SU(2)_L \times
U(1)_Y$ symmetry is spontaneously broken down to $\Uoneem$.
Accordingly, the field $\Bphi$ describes one massive mode, the Higgs
particle, and three Goldstone bosons which render the gauge fields $W$
and $Z$ massive. Finally, the spectrum contains the massless photon.
At tree level, the masses and the electric coupling constant $e$ are
given by the relations  
\be \label{bare_masses_couplings}
M_H^2 = 2 m^2 \, , \, 
M_W^2 = {m^2 g^2 \over 4 \lambda} \, , \, 
M_Z^2 = {m^2 (g^2 + \gp^2) \over 4 \lambda} \, , e^2 = {g^2 \gp^2
  \over g^2 + \gp^2} \ . 
\ee
Furthermore we will use the following definition of the weak mixing
angle:  
\be \label{cos_theta}
c^2 \equiv \cos^2 \theta_W = M_W^2 / M_Z^2 \, , \, 
s^2 \equiv 1 - c^2 \ . 
\ee

In order to have nontrivial solutions of the equations of motion, we
furthermore couple external sources to the gauge fields and the Higgs
boson. As discussed in detail for the Abelian Higgs model in
Ref.~\cite{Abelian_Higgs} and for QED in Ref.~\cite{QED_gaugeinv}, the
appropriate choice of the source terms is crucial for a manifestly
gauge-invariant analysis.

The sources will only respect the gauge symmetry, if they do not
couple to the gauge degrees of freedom. Otherwise, one has to impose
constraints on the fields in order to solve the equations of motion.
Usually, this problem is cured by fixing a gauge. However, one can
also turn the argument around and consider only those external sources
which couple to gauge-invariant operators. As we will see below, such
a manifestly gauge-invariant treatment is in fact possible at the
classical level as well as when quantum corrections are taken into
account.

In order to write down appropriate source terms we will introduce
another set of fields for the dynamical degrees of freedom which are
already invariant under the non-Abelian group $SU(2)_L$ and, in parts,
under the Abelian group $U(1)_Y$ as well. It has been known for a long
time \cite{EW_confinement,tHooft,gauge_inv_fields} that all fields
in the standard model Lagrangian can be written, in the spontaneously
broken phase, in a gauge-invariant way up to the unbroken
$\Uoneem$. It is convenient to use a polar representation for the
Higgs doublet field
\be \label{polar}
\Bphi = {m\over \sqrt{\lambda}} R U \, , 
\ee
where the unitary field~$U$, satisfying~$U^\dagger U = 1$, describes
the three Goldstone bosons, while the radial component~$R$ represents
the Higgs boson. Furthermore, we define the $Y$-charge conjugate doublet 
\be \label{tildePhi} 
\tBphi = i \tau_2 \Bphi^* \, ,   
\ee
and similarly, $\tU = i \tau_2 U^*$. 

We introduce the following operators: 
\bea 
V_\mu^1 & = & i \tBphi^\dagger D_\mu\Bphi + i \Bphi^\dagger
D_\mu\tBphi  
= {m^2\over\lambda} R^2 {\cal W}_\mu^1 \, , \nonumber \\
V_\mu^2 & = & - \tBphi^\dagger D_\mu\Bphi +  \Bphi^\dagger D_\mu\tBphi 
= {m^2\over\lambda} R^2 {\cal W}_\mu^2 \, , \nonumber \\
V_\mu^3 & = & i \tBphi^\dagger D_\mu\tBphi - i \Bphi^\dagger D_\mu\Bphi 
= {m^2\over\lambda} R^2 {\cal Z}_\mu \, , \label{bcomposite}
\eea
involving the gauge boson fields 
\begin{eqnarray}\label{bfieldsf}
{\cal W}^+_\mu &=& {i\over2} \left(\tilde U^\dagger (D_\mu U) -
(D_\mu\tilde U)^ 
\dagger U\right) \, , \\
{\cal W}^-_\mu &=& {i\over2} \left( U^\dagger (D_\mu \tilde U) -
(D_\mu U)^\dagger \tilde U\right) \, , \\
{\cal Z}_\mu &=& i \left( \tilde U^\dagger (D_\mu \tilde U) -
U^\dagger (D_\mu U) \right) \, , \label{defZ} \\
{\cal A}_\mu &=& B_\mu + s^2 {\cal Z}_\mu \, , \label{bfieldsl} \\
{\cal W}^\pm_\mu &=& {1\over2} ({\cal W}^1_\mu \mp i {\cal W}^2_\mu)
\, , 
\end{eqnarray}
which are invariant under the $SU(2)_L$ gauge transformations from
Eq.~(\ref{SUtwo_trafo}). Up to a constant factor the
operators~$V_\mu^i$ in Eq.~(\ref{bcomposite}) correspond to the
currents of the global symmetry~$SU(2)_R$.

In terms of these composite fields the Lagrangian from
Eq.~(\ref{Lag_SM_before}) reads    
\be
\label{Lag_SM_inv}
\lag_\SM^0 = {1\over 2} {m^2\over \lambda}
\left[ \p_\mu R \p_\mu R - m^2 R^2 + {m^2\over 2} R^4 + R^2
\left(\W_\mu^+ \W_\mu^- + {1\over 4} \Z_\mu \Z_\mu \right) \right]
+ {1\over 4\gs} \W_{\mu\nu}^a \W_{\mu\nu}^a
+ {1\over 4\gps} \B_{\mu\nu} \B_{\mu\nu} \, , 
\ee
where
\bea
\W_{\mu\nu}^a   & = & \p_\mu \W_\nu^a - \p_\nu \W_\mu^a
                        + \varepsilon^{abc} \W_\mu^b \W_\nu^c \ , 
                        \ a = 1,2,3 \, , \\
\W_\mu^3        & = & \Z_\mu + B_\mu \ . 
\eea

In order to calculate Green's functions from which we then can extract
physical quantities like masses, coupling constants and $S$-matrix
elements, we have to introduce external sources which emit
one-particle states of the Higgs field and the gauge bosons. In
analogy to the Abelian case~\cite{Abelian_Higgs,QED_gaugeinv} we
couple sources to the $SU(2)_L \times U(1)_Y$ gauge-invariant operator
$\Bphi^\dagger \Bphi$ and the field strength $B_{\mu\nu}$. For the
massive gauge bosons the situation is more involved. Whereas the field
$\Z_\mu$ is fully gauge-invariant, the charged gauge fields
$\W_\mu^\pm$ and the corresponding currents $V_\mu^\pm$ have a
residual gauge dependence under the $U(1)_Y$ gauge transformations
from Eq.~(\ref{Uone_trafo}):
\be 
\W_\mu^{\pm} \to e^{\mp i \UoneParam} \W_\mu^{\pm} \quad , \quad 
V_\mu^{\pm} \to e^{\mp i \UoneParam} V_\mu^{\pm} \ . 
\label{calW_trafo} 
\ee
We can, however, compensate this gauge dependence by multiplying the
charged fields $\W_\mu^\pm$ and $V_\mu^\pm$ by a phase
factor~\cite{Steinmann,Horan_Lavelle_McMullan,QED_gaugeinv}. In
terms of the operators $V_\mu^a$ from Eq.~(\ref{bcomposite}) we can
then write appropriate $SU(2)_L \times U(1)_Y$ gauge-invariant source
terms for all the fields as follows:
\be
\label{basic_sources}
\lag_{\text{source}}^1 =  
- \half h \Bphi^\dagger \Bphi   
- \half K_{\mu\nu} B_{\mu\nu}  
+ J_\mu^a \phase^{ab} V_\mu^b \, ,  
\ee
with external sources $h, K_{\mu\nu}$, and $J_\mu^a (a=1,2,3)$. The
phase factor in Eq.~(\ref{basic_sources}) is defined by 
\be \label{phase_matrix}
\phase(x)     = \exp\left({T \int d^dy \, \G_0(x-y) \, 
\partial_\mu \B_\mu(y)}\right) \, , 
\ee
with
\be \label{matrix_T}
T  =  \left( 
\begin{array}{ccc} 
0  & 1 & 0 \\
-1 & 0 & 0 \\
0  & 0 & 0
\end{array}
\right) \, , 
\ee
and
\be
\G_0(x-y) = \langle x | {1 \over - \Box} | y \rangle . 
\ee
Since the vacuum in the spontaneously broken phase corresponds to the
value~$R=1$, Green's functions of the field $\Bphi^\dagger \Bphi$
contain one-particle poles of the Higgs boson, whereas those of
$\phase^{ab} V_\mu^b$ have one-particle poles of the gauge bosons $W$ 
and $Z$. 

In Ref.~\cite{QED_gaugeinv} it was shown to all orders in perturbation
theory that a phase factor $\phase$ which is defined analogously to
Eq.~(\ref{phase_matrix}) does not spoil the renormalizability of QED.
Since the proof did not rely on any particular feature of QED, the
same should be true for the present case as well. This is due to the
fact that the phase factor only contains the Abelian gauge degree of
freedom which does not affect the dynamics of the theory. Since the
operator $\Bphi^\dagger \Bphi$ and the currents $V_\mu^a$ from
Eq.~(\ref{bcomposite}) have dimension less than four, source terms
involving these operators do not spoil the renormalizability
either. The reader should note, however, that we do not have a formal
proof of renormalizability to all orders in perturbation theory for
the present case. As will be shown below, at the one-loop level
everything works fine and on physical grounds we expect this to happen
at all orders.

Green's functions of the operators in Eq.~(\ref{basic_sources}) are,
however, more singular at short distances than (gauge-dependent)
Green's functions of the fields $\Bphi, W_\mu^a,$ and $B_\mu$
themselves.  Time ordering of these operators gives rise to
ambiguities, and the corresponding Green's functions are only unique
up to contact terms. In order to make the theory finite, these contact
terms of dimension four need to be added to the Lagrangian which is
then given by 
\be \label{lagSMfull} 
\lag_\SM = \lag_\SM^0 + \widehat\lag_{\text{source}}^1 +
\lag_{\text{source}}^2 \, .  
\ee
The first term in Eq.~(\ref{lagSMfull}) is defined in
Eq.~(\ref{Lag_SM_inv}). The second term is given by 
\be
\label{lag_sources_1_hat}
\widehat\lag_{\text{source}}^1 =  
- \half \widehat h \Bphi^\dagger \Bphi   
- \half \widehat K_{\mu\nu} B_{\mu\nu}  
+ J_\mu^a \phase^{ab} V_\mu^b \, , 
\ee
where
\bea
\widehat h  & = & h + 4 v_{jj} J_{\mu}^+ J_{\mu}^-
        + c_{jj} J_{\mu}^\Z J_{\mu}^\Z + 4 J_\mu^a J_\mu^a \, , \\
\widehat K_{\mu\nu} & = & K_{\mu\nu} + c_{Bj} \jmunut
        - 2 i c_{Bjj} (J_\mu^+ J_\nu^- - J_\mu^- J_\nu^+) \, . 
\eea
The last term in Eq.~(\ref{lagSMfull}) is defined by 
\bea
\lag_{\text{source}}^2 & = &\mbox{}- v_{djj} J_{\nu}^\Z [ i (d_\mu
J_{\nu}^+ - d_\nu J_{\mu}^+) 
J_\mu^- - i (d_\mu J_{\nu}^- - d_\nu J_{\mu}^-) J_\mu^+] 
+v_{dj} (d_\mu J_{\nu}^+ - d_\nu J_{\mu}^+)  (d_\mu
J_{\nu}^- - d_\nu J_{\mu}^-) \nonumber \\
&&\mbox{}- {i\over 2} c_{djj} \jmunut (J_\mu^+ J_\nu^- - J_\mu^-
J_\nu^+) +{1\over4} c_{dj} \jmunut \jmunut \nonumber\\
&&\mbox{}+ 16 v_{JJ2} (J_{\mu}^+ J_{\mu}^-)^2
+ 4 v_{JJJJ} (J_{\mu}^+ J_{\nu}^- + J_\mu^- J_\nu^+)^2
+ c_{JJ2} (J_{\mu}^\Z J_{\mu}^\Z )^2 \nonumber\\
&&\mbox{}+ 4 v_{J2ZZ} J_{\mu}^+ J_{\mu}^- J_{\nu}^\Z J_{\nu}^\Z
+ 2 v_{JJZZ} (J_{\mu}^+ J_{\nu}^- + J_\mu^- J_\nu^+) J_{\mu}^\Z
J_{\nu}^\Z  \nonumber\\ 
&&\mbox{}+ c_{hh} h^2 + c_{mh} m^2 h 
+ 4 c_{hJJ} h  J_{\mu}^+ J_{\mu}^-
+ 4 c_{mJJ} m^2  J_{\mu}^+ J_{\mu}^- 
+ c_{hZZ} h J_{\mu}^\Z J_{\mu}^\Z
+ c_{mZZ} m^2 J_{\mu}^\Z J_{\mu}^\Z  \, , 
\label{lagsourcetwo}
\eea
where we introduced the quantities 
\bea
J_\mu^\pm       & = & {1\over 2} (J_\mu^1 \mp i J_\mu^2) \ , \quad
J_\mu^\Z        \equiv  J_\mu^3 \, , \\
d_\mu J_\nu^\pm & = & (\p_\mu \mp i B_\mu^T) J_\nu^\pm \ , \quad 
B_\mu^T = \PT_{\mu\nu} B_\nu \, , \label{dmuJnupm} \\
\PT_{\mu\nu}     & = & \delta_{\mu\nu} - \PL_{\mu\nu} \, , \quad 
\PL_{\mu\nu} = {\p_\mu \p_\nu \over \Box} \ . 
\eea
The contact terms in $\lag_{\text{source}}^2$ will not contribute to any
physical $S$-matrix elements. 

For later use we define the following $SU(2)_L \times U(1)_Y$
gauge-invariant fields:  
\bea
\u V_\mu^a      & = & \phase^{ab} V_\mu^b \, ,  \label{def_uV} \\
\u\W_\mu^\pm      & = & \phase^\mp \W_\mu^\pm \, ,  \label{def_uW} \\
\A_\mu^T          & = & \PT_{\mu\nu} \A_\nu \, ,    \label{def_AT}  
\eea
where 
\be \label{phase_complex}
\phase^\mp(x)      =  \exp\left({\pm i \int d^dy  \, \G_0(x-y) \,
\partial_\mu \B_\mu}(y) \right)  \ . 
\ee
The projection on the transverse mode in Eq.~(\ref{def_AT}) leads to a
fully $SU(2)_L \times U(1)_Y$ gauge-invariant field, since the
$SU(2)_L$ invariant field $\A_\mu$ from Eq. (\ref{bfieldsl})
transforms under $U(1)_Y$ as follows:
\be
\A_\mu \to  \A_\mu - \partial_\mu \UoneParam \, , 
\ee
i.e.\ like an Abelian gauge field. 

Furthermore we introduce the quantities
\bea
\Y_\mu^\pm & = & \W_\mu^\pm + 4 j_\mu^\pm \ , \ 
\Y_\mu^\Z  =  \Z_\mu + 4 J_\mu^\Z  \, , \\
j_\mu^\pm      & = & \phase^\pm J_\mu^\pm \ . 
\eea

The generating functional $W_\SM[h,K_{\mu\nu},J_\mu^a]$ for the
gauge-invariant Green's functions is defined by the path integral
\begin{equation} \label{genfunc_pathint}
   e^{-W_\SM [h,K_{\mu\nu},J_\mu^a]} = \int \d\mu[\Bphi,W_\mu^a,B_\mu]
        e^{-\intdx \lag_\SM} \ . 
\end{equation}
Note that we still integrate over the original fields $\Bphi,
W_\mu^a,$ and $B_\mu$ in Eq. (\ref{genfunc_pathint}). Furthermore, we
have absorbed an appropriate normalization factor into the measure
$\d\mu[\Bphi,W_\mu^a,B_\mu]$. Derivatives of this functional with
respect to the field $h$ generate Green's functions of the scalar
density $\Bphi^\dagger\Bphi$, derivatives with respect to the source
$K_{\mu\nu}$ generate Green's functions of the field strength
$B_{\mu\nu}$, while derivatives with respect to $J_\mu^a$ generate
Green's functions for the currents $\u V_\mu^a$.

In the spontaneously broken phase, these Green's functions have
one-particle poles from the Higgs boson as well as the gauge bosons.
Thus, one can extract $S$-matrix elements for the physical degrees of
freedom from the generating functional in Eq.~(\ref{genfunc_pathint}).
Due to the equivalence theorem~\cite{equivalence_theorem} these
$S$-matrix elements will be identical to the ones obtained from those
Green's functions which are used in the usually employed formalism.
The presence of the contact terms in $\lag_{\text{source}}^2$ in
Eq.~(\ref{lagsourcetwo}) reflects the fact that the off-shell
continuation of the $S$-matrix is not unambiguously defined. Note that
this is a general feature of any field theory and not particular to
those involving a gauged symmetry. The continuation we choose here has
the virtue of being gauge-invariant.

There is another aspect worth noting. In
Refs.~\cite{EW_confinement,tHooft} it was pointed out that the
complete screening of the $SU(2)_L$ charge of the composite fields
$V_\mu^a$ and $\Bphi^\dagger \Bphi$ can be interpreted as the
manifestation of confinement in the electroweak theory, similarly to
the mechanism in QCD. As discussed in Ref.~\cite{tHooft} the
physically observed particles then correspond to ``mesonic'' and
``baryonic'' bound states of the usual fields that appear in the
Lagrangian. To illustrate this point more clearly, it is useful to
include the fermions for a moment. As shown in
Ref.~\cite{QED_gaugeinv}, for up and down type quarks and leptons one
may consider the following composite fields:
\bea 
\Bphi^\dagger q_L^k & = & {m\over \sqrt{\lambda}} R d_L^k  , \qquad 
\tBphi^\dagger q_L^k = {m\over \sqrt{\lambda}} R u_L^k  , \nonumber \\
\Bphi^\dagger l_L^k & = & {m\over \sqrt{\lambda}} R e_L^k  , \qquad 
\tBphi^\dagger l_L^k = {m\over \sqrt{\lambda}} R \nu_L^k  ,
\label{Higgs_Fermion} 
\eea
which appear in the Yukawa interactions. The interpolating fermion
fields  
\bea
d_L^k & = & U^\dagger q_L^k , \qquad 
u_L^k = \tU^\dagger q_L^k , \nonumber \\ 
e_L^k & = & U^\dagger l_L^k , \qquad  
\nu_L^k = \tU^\dagger l_L^k ,  \label{def_u_d_e_nu}
\eea
are $SU(2)_L$-invariant and have the same $U(1)_Y$ quantum numbers as
their right-handed counterparts. The fields $q_L^k$ and $l_L^k$ in
Eqs.~(\ref{Higgs_Fermion}) and (\ref{def_u_d_e_nu}) are the usual
fermion doublets for the quarks and leptons with family index $k$.

Denoting schematically all doublets (Higgs, quarks, leptons) by $q$,
one can interpret the fields built out of $q^\dagger q^\prime$ as
``mesons'' and the fields $\tilde q^\dagger q^\prime \equiv - q_i
\epsilon^{ij} q_j^\prime$ as ``baryons,'' cf.\ Eq.~(\ref{tildePhi}). The
fields $q^\dagger D_\mu q^\prime$ and $\tilde q^\dagger D_\mu
q^\prime$ can then be viewed at as $P$-wave states of these ``mesons''
and ``baryons.'' Thus we have the following ``mesons'' $R d_L^k, R
e_L^k, R^2 \Z_\mu$ and $\Bphi^\dagger \Bphi,$ which correspond for
each family $k$ to the physical $d$-quark\footnote{Of course, if we
switch on the QCD interactions, the quarks will be confined in
hadrons.}, the electron, the $Z$-boson and to the Higgs boson,
respectively. Furthermore, there are the following ``baryons'' $R
u_L^k, R \nu_L^k$ and $R^2 \W_\mu^\pm,$ which correspond to the
physical $u$-quark, the neutrino and the $W$-bosons. The fundamental
fields $l_L^k, q_L^k$ and $\Bphi$ which carry $SU(2)_L$-charges are
confined at low energies, i.e.\ around the electroweak scale, in these
``mesons'' and ``baryons'' due to the strong non-Abelian forces of the
$SU(2)_L$ gauge fields. Therefore the $SU(2)_L$ charge cannot be
observed in physical states, similarly to color in QCD. Note that the
notions ``meson'' and ``baryon'' are convention dependent. In
particular, we use different conventions than those employed in
Ref.~\cite{tHooft}.

Our approach, extending the gauge-invariant treatment to the full
group $SU(2)_L \times U(1)_Y$, can thus be viewed at as a well-defined
framework for carrying out calculations which involve only those
external fields which correspond to the physically observed particles.

As was pointed out in Refs.~\cite{Abelian_Higgs,QED_gaugeinv} it is
possible to evaluate the path integral in Eq.~(\ref{genfunc_pathint})
without the need to fix a gauge as will be shown in the following.


\section{Tree level}
\label{sec:tree_level}

At tree level, the generating functional for the bosonic sector of the 
standard model is given by 
\begin{equation} \label{genfunc_tree}
        W_\SM[h,K_{\mu\nu},J_\mu^a] 
                = \intdx \, \lag_\SM(R^{cl}, \u\W_\mu^{cl,\pm},
                \Z_\mu^{cl}, \A_\mu^{cl,T}) \ , 
\end{equation}
where $R^{cl}, \u\W_\mu^{cl,\pm}, \Z_\mu^{cl},$ and $\A_\mu^{cl,T}$
are determined by the equations of motion 
\bea
- \Box R & = & - \left[ m^2 (R^2 - 1) + \u\Y_\mu^+ \u\Y_\mu^- +
{1\over4} \Y_\mu^\Z \Y_\mu^\Z - \widehat h  \right] R \, ,
\label{eomRcomp} \\ 
- \u d_\mu \u\W_{\mu\nu}^\pm & = & - M_W^2 R^2 \left( \u\W_\nu^\pm + 4
J_\nu^\pm \right) \pm i \left( c^2 \Z_{\mu\nu} + \A_{\mu\nu} \right)
\u\W_\mu^\pm 
\pm 2 \left( \u\W_\mu^+ \u\W_\nu^- - \u\W_\mu^-
\u\W_\nu^+ \right) \u\W_\mu^\pm \, , \label{eomWcomppm} \\
-\p_\mu\Z_{\mu\nu} & = & \PT_{\nu\mu}\left(-M_Z^2 R^2 \Y_\mu^\Z +
\DeltaEoM_\mu\right) + {e^2\over c^2} \p_\mu \widehat K_{\mu\nu} 
+ {e^2\over c^2} \PT_{\nu\mu} S_\mu \, , \label{eomZcomp} \\
-\p_\mu\A_{\mu\nu} &=& s^2 \PT_{\nu\mu}\DeltaEoM_\mu - e^2 \p_\mu
\widehat K_{\mu\nu} - e^2 \PT_{\nu\mu} S_\mu \ . \label{eomAcomp} 
\eea
Furthermore, the equations for the Goldstone boson field $U$
correspond to 
\bea
\u d_\mu \u\Y_\mu^\pm & = & - 2 {\p_\mu R \over R} \u\Y_\mu^\pm \mp i
\y_\mu \u\W_\mu^\pm \, , \label{eomUcomppm} \\
\p_\mu\y_\mu &=& - 2 {\p_\mu R\over R} \y_\mu - 8 i (
J^+_\mu\u\W^-_\mu - J_\mu^-\u\W^+_\mu) \ . 
\label{eomUcompZ} 
\eea
In order to simplify the notation we have omitted the prescription 
``cl'' in the equations above. In
Eqs.~(\ref{eomRcomp})--(\ref{eomUcompZ}) we have introduced the
quantities  
\bea
\u\Y_\mu^\pm       & = & \phase^\mp \Y_\mu^\pm 
= \u\W_\mu^\pm + 4 J_\mu^\pm \, , \\
\u d_\mu \u\W_\nu^\pm & = & \left( \p_\mu \mp i \left[ \Z_\mu 
+ s^2 \Z_\mu^T - \A_\mu^T \right] \right) \u\W_\nu^\pm \, , 
\label{ud_uWpm} \\
\u\W_{\mu\nu}^\pm & = & \u d_\mu \u\W_\nu^\pm - \u d_\nu \u\W_\mu^\pm
\, , \\
\Z_{\mu\nu} & = & \p_\mu \Z_\nu - \p_\nu \Z_\mu \, , \\
\A_{\mu\nu} & = & \p_\mu \A_\nu - \p_\nu \A_\mu \, , \\
\DeltaEoM_\mu &=&  2 \Z_\rho ( \u\W_\rho^+ \u\W_\mu^- + \u\W_\mu^+
\u\W_\rho^- ) - 4 \Z_\mu \u\W_\rho^+ \u\W_\rho^- 
+ 2 i ( \u\W^+_{\rho\mu} \u\W^-_\rho  -
\u\W^-_{\rho\mu} \u\W^+_\rho ) \nonumber \\ 
&&\mbox{}- 2 i ( \u d_\rho\u\W_\rho^+\u\W_\mu^-  - 
\u d_\rho\u\W_\rho^-\u\W_\mu^+  
- \u d_\rho\u\W_\mu^+\u\W_\rho^- + \u d_\rho\u\W_\mu^-\u\W_\rho^+) 
\, , \\
S_\mu   & = &\mbox{}- v_{djj} J^\Z_{\rho} (J_{\rho}^+ J_{\mu}^- +
J_\rho^- J_\mu^+) + 2 v_{djj} J^\Z_{\mu} J_{\rho}^+ J_{\rho}^-
        \nonumber \\
        &&\mbox{}- 2 v_{dj} [i (d_\rho J_{\mu}^+ - d_\mu J_{\rho}^+)
        J_\rho^- - i (d_\rho J_{\mu}^- - d_\mu J_{\rho}^-) J_\rho^+] 
\ . 
\eea
The covariant derivatives in $d_\mu J_\nu^\pm, \u d_\mu \u\Y_\nu^\pm,$
and $\u d_\mu \u \W_{\mu\nu}^\pm$ are defined in the same way as in
Eq.~(\ref{ud_uWpm}). 

Several things about Eqs.~(\ref{eomRcomp})--(\ref{eomUcompZ})
are worth being noticed. First of all, only the physical degrees of
freedom enter these equations. The radial variable $R$ which is
related to the massive Higgs boson is determined by
Eq.~(\ref{eomRcomp}). Solutions for the massive gauge boson fields
$\u\W_\mu^\pm$, cf.\ Eq.~(\ref{def_uW}), and $\Z_\mu$ follow from
Eqs.~(\ref{eomWcomppm}) and (\ref{eomZcomp}).  Finally,
Eq.~(\ref{eomAcomp}) determines the transverse component of the
massless photon field $\A_\mu^T$.  Note that the equations of motion
do not determine the longitudinal component of the photon field and
the phase of the gauge boson fields $\W_\mu^\pm$ which correspond to
the $U(1)_Y$ gauge degree of freedom. Even more they do not determine
the classical Goldstone boson field $U$ either, since it corresponds
to the $SU(2)_L$ gauge degrees of freedom. Thus, gauge invariance
implies that these equations have a whole class of solutions in terms
of the original fields $\Bphi, W_\mu^a, B_\mu$.  Every two
representatives are related to each other by a gauge transformation.
Nevertheless, the physical degrees of freedom are uniquely determined
by these equations of motion. Moreover, since the action is
gauge-invariant, the generating functional in Eq.~(\ref{genfunc_tree})
is uniquely determined as well.

The most important point is the fact that the classical Goldstone
boson field $U$ represents the $SU(2)_L$ gauge degrees of freedom.
Thus, no Goldstone bosons are propagating at the classical level of
the theory. All gauge-invariant sources emit physical modes only.
Moreover, Eqs.~(\ref{eomUcomppm}) and (\ref{eomUcompZ}), which follow
from the requirement that the variation of the Lagrangian with respect
to the Goldstone boson field $U$ vanishes, are not equations of
motion, but constraints expressing the fact that the gauge fields
$\u\W_\mu^\pm, \Z_\mu,$ and $\A_\mu$ couple to conserved currents.
They can also be obtained by taking the derivative of the equations of
motion for the gauge fields. Note that we have already used the
constraints to bring these equations of motion into the form given in
Eqs.~(\ref{eomRcomp})--(\ref{eomAcomp}).

In order to solve the classical equations of
motion~(\ref{eomRcomp})--(\ref{eomAcomp}) we introduce a parameter
$\source$ which counts powers of the external sources:
\be
h, K_{\mu\nu}, J_\mu^a  = \order(\source) \ . 
\ee
From this we get the counting rules 
\be
R-1, \u\W_\mu^\pm, \Z_\mu, \A_\mu =  \order(\source)
\ . 
\ee
We will see below that this counting scheme is self-consistent. 

The solution of the equation of motion for the Higgs field $R$,
Eq.~(\ref{eomRcomp}), reads, up to and including quadratic terms in 
powers of the external sources,  
\bea 
(R-1)(x) & = & \int d^dy \, \G_H(x-y) \left( \widehat h  - \Y_\mu^+
  \Y_\mu^- - {1\over4} \Y_\mu^\Z \Y_\mu^\Z \right)(y) 
+ \int d^dy d^dz \, \G_H(x-y) \left( h(y) \G_H(y-z) h(z) \right) 
\nonumber \\
& &\mbox{}- 3m^2 \int d^dy \, \G_H(x-y)  \left( \int d^dz \, \G_H(y-z) h(z)
\right)^2 \, . \label{solution_R}
\eea
In order to calculate the two-point functions of the physical 
fields in Sec.~\ref{sec:greensfunctions} we will not need the terms
of third and higher order in powers of the external sources in
Eq.~(\ref{solution_R}).  

The solutions for the equations of motion for the gauge fields,
Eqs.~(\ref{eomWcomppm})--(\ref{eomAcomp}), are given by 
\bea
\u\W_\mu^{\pm,T}(x)  & = & \int d^dy \G_W(x-y) \left( - 4 M_W^2
J_\mu^{\pm,T} \right)(y) \, , 
  \label{solution_Wpm} \\
\u\W_\mu^{\pm,L}(x)  & = & - 4 J_\mu^{\pm,L}(x) \, , \\ 
\Z_\mu^T(x)        & = & \int d^dy \G_Z(x-y)\left( - 4 M_Z^2
  J_\mu^{\Z,T} + {e^2\over c^2} \p_\rho \widehat K_{\rho\mu}
  \right)(y) \, , \\ 
\Z_\mu^L(x)        & = & - 4 J_\mu^{\Z,L}(x) \, , \\ 
\A_\mu^T(x)        & = & \int d^dy \G_A(x-y) \left( - e^2 \p_\rho \widehat
  K_{\rho\mu} \right)(y)  \ . 
\label{solution_AT} 
\eea
We will only need the leading terms of the solution in powers of the
external sources later on. In
Eqs.~(\ref{solution_R})--(\ref{solution_AT}) we have introduced the
quantities  
\bea
J_\mu^{\pm,T} & = & \PT_{\mu\nu} J_\nu^\pm \, , \qquad J_\mu^{\Z,T} =  
\PT_{\mu\nu} J_\nu^\Z \, , \nonumber \\ 
J_\mu^{\pm,L} & = & \PL_{\mu\nu} J_\nu^\pm \, , \qquad J_\mu^{\Z,L} = 
\PL_{\mu\nu} J_\nu^\Z \, , 
\eea
and 
\bea
\G_H(x-y) & = & \langle x | {1\over -\Box + 2m^2}  | y \rangle \, ,
\nonumber \\ 
\G_W(x-y) & = & \langle x | {1\over -\Box + M_W^2} | y \rangle \, ,
\nonumber \\ 
\G_Z(x-y) & = & \langle x | {1\over -\Box + M_Z^2} | y \rangle \, ,
\nonumber \\ 
\G_A(x-y) & = & \langle x | {1\over -\Box} | y \rangle \, . 
\label{treelevel_prop}
\eea

The $U(1)_Y$ gauge degree of freedom of the longitudinal component of
$B_\mu^L$ can be parametrized as follows:
\be
B_\mu^L = \partial_\mu \gaugeparam \, , 
\ee
with an arbitrary function $\gaugeparam$. The solutions for the
longitudinal component of the photon $\A_\mu^L$ and the phase of the
$W$-boson field $\W_\mu^\pm$ are given by 
\bea
\A_\mu^L & = & \partial_\mu \gaugeparam + s^2 \Z_\mu^L \, , \\ 
\W_\mu^\pm & = & e^{\pm i \gaugeparam} \u\W_\mu^\pm \ . 
\eea
The gauge dependence of these fields manifests itself through the
presence of the undetermined function $\gaugeparam$.  The $SU(2)_L$
gauge invariance corresponds to the freedom to choose an arbitrary
field $U$. For instance, the choice $U = {0 \choose 1}$ leads to the
unitary gauge.


\section{One-loop level}
\label{sec:one_loop}

The one-loop contribution to the generating functional can be
evaluated with the saddle-point method. Before we proceed with the
explicit calculation of the generating functional some general remarks
are in order.  If we write the fluctuations $y$ around the classical
fields $\Allfields^{cl}$ as $\Allfields = \cAllfields + y$, we obtain
the following representation for the one-loop approximation to the
generating functional: 
\begin{equation} \label{genfunc_saddlepoint}
  e^{-W_\SM[h,K_{\mu\nu},J_\mu^a]} = e^{-\intdx\lag_\SM^{cl}} 
     \int\d\mu[y] e^{- (1/2) \intdx y^T \tildeD y} \ .
\end{equation}
Gauge invariance implies that the operator $\tildeD$ has zero
eigenvalues corresponding to fluctuations $y$ which are equivalent to
infinitesimal gauge transformations. Indeed, if $\Allfields^{cl,i}$ is a
solution of the equation of motion, i.e., a stationary point of the
classical action,
\begin{equation} \label{geqmo}
        \left.{\delta
          S_\SM\over\delta\Allfields^i}\right|_{\Allfields=\cAllfields} = 0
        \ , 
\end{equation}
then any gauge transformation yields another equivalent solution. The
index $i$ in $\Allfields^{cl,i}$ labels the different fields. 
Thus, differentiating equation~(\ref{geqmo}) with respect to the gauge
parameters $\gaugeparam^A$ one obtains 
\begin{equation} \label{zero1}
        \left.
        {\delta^2 S_\SM\over\delta\Allfields^i\delta\Allfields^j}
        {\delta\Allfields^j\over\delta\gaugeparam^A}
        \right|_{\Allfields=\cAllfields} = 0 \ .
\end{equation}
The quadratic form which appears in Eq.~(\ref{zero1}) is identical to
the differential operator $\tildeD$. If we denote the zero eigenvector
by $\zeta$ and parametrize it in terms of scalar fields $\alpha$ by way of
$\zeta = P\alpha$, with some differential operator $P$,
Eq.~(\ref{zero1}) translates to the identity $P^T \tildeD  =  \tildeD
P  = 0$.  
Let $\alpha_m$ be the eigenvectors of the operator $P^T P$,
i.e. 
\begin{equation}
        P^T P \alpha_m = l_m \alpha_m \ .
\end{equation}  
Then, the expansion of the fluctuation $y$ in terms of eigenvectors of the
operator $\tildeD$ is given by
\begin{equation}
        y = \sum_n a_n \xi_n + \sum_m b_m \zeta_m \ ,
\end{equation}
where $\zeta_m = P \alpha_m$ and $\xi_n$ have zero and non-zero eigenvalues,
respectively.

In order to evaluate the path integral in
Eq.~(\ref{genfunc_saddlepoint}), we use Polyakov's
method~\cite{Polyakov} and equip the space of fields with a metric 
\be
    ||y||^2 = \intdx y^T y = \sum_n a_n^2 + \sum_m b_m^2 l_m \ .
\ee
With our choice for the scalar fields $\alpha_m$, the metric on the
kernel of the differential operator $\tildeD$ is diagonal:
\begin{equation}
  g_{\bar m m} = \intdx \alpha_{\bar m} P^T P \alpha_m = \delta_{\bar
    m m} l_m \, ,    
\end{equation}
and the volume element associated with this metric is then given by
\begin{equation}
   \d\mu[\Bphi, W_\mu^a ,B_\mu] = {\cal N} \prod_n \d a_n \prod_m \d b_m
   \sqrt{ \det P^T P } \ .  
\end{equation}
The integration over the zero modes yields the volume factor of the
gauge group, which can be absorbed by the normalization of the
integral. The remaining integral over the non-zero modes is damped by
the usual Gaussian factor. Up to an irrelevant infinite constant one
obtains the following result for the one-loop generating
functional from Eq.~(\ref{genfunc_saddlepoint}):  
\begin{equation} \label{1loopgf}
  W_\SM[h,K_{\mu\nu},J_\mu^a] = \intdx \lag_\SM 
  + \half {\ln\det}^\prime\tildeD
  -\half \ln\det P^T P \ .
\end{equation}
The first term on the right-hand side represents the classical action
which describes the tree-level contributions to the generating
functional. In the second term, the determinant ${\det}^\prime\tildeD$
is defined as the product of all non-zero eigenvalues of the operator
$\tildeD$. The last term originates from the path integral measure.
The sum of the last two terms in Eq.~(\ref{1loopgf}) corresponds to
the one-loop contributions to the generating functional.

We now discuss in more detail the evaluation of the one-loop
contributions to the generating functional in
Eq.~(\ref{genfunc_saddlepoint}) for the standard model. The choice of
an appropriate parametrization of the physical modes and their quantum
fluctuations is very important in order to obtain an expression for
the differential operator which is still tractable. We introduce the
following fluctuations $f,\eta^a,\w_\mu^a,$ and $\b_\mu$ around the
Higgs field $R$, the Goldstone boson field $U$, the three $SU(2)_L$
gauge fields $W_\mu^a$ and the $U(1)_Y$ gauge field $B_\mu$,
respectively:
\begin{eqnarray}
R       & \rightarrow & R + {\sqrt{\lambda}\over m} f \, ,
\label{fluct_R} \\ 
U       & \rightarrow & e^{i \kappa /2} V U \, , \label{fluct_U}
\\ 
W_\mu  & \rightarrow & W_\mu + {1\over
  2} \g \w_\mu^a V t^a V^\dagger , \quad W_\mu \equiv W_\mu^a
{\tau^a\over 2} \, , \\ 
B_\mu   & \rightarrow & B_\mu  + \gp \b_\mu \, , \label{fluct_B}
\eea
where 
\bea
\kappa(x)  & = &  - g^\prime \intdy \G_0(x-y) \p_\mu b_\mu(y) \, , \\
V(x)       & = & \exp \left( i {\sqrt{\lambda}\over m R(x)} \eta^a(x)
t^a(x) \right)  , \quad V \in SU(2) \ . \label{def_V} 
\eea
The matrices $t^i$ in Eq.~(\ref{def_V}) are defined through the
relations 
\bea 
t^1     & = & U \tU^\dagger + \tU U^\dagger \, , \nonumber \\
t^2     & = & i (U \tU^\dagger - \tU U^\dagger) \, , \nonumber \\
t^3     & = & \tU \tU^\dagger - U U^\dagger \, , 
\eea
and satisfy the Pauli algebra
\be
\left[ t^i , t^j \right]  = 2 i \varepsilon^{ijk} t^k  \ , \quad 
\{ t^i , t^j \} =  2 \delta^{ij} \mathbf{1} \ .  
\ee
Using the above transformation properties we get 
\bea
\W_\mu^a & \rightarrow & \left(e^{T\kappa}\right)^{ab}\left(
\W_\mu^b + \g \w_\mu^b 
        + g^\prime \PL_{\mu\nu} \b_\nu \delta^{b3}
+ i \ \tr (t^b V^\dagger D_\mu V) \right) \nonumber \\
        & = & \left(e^{T\kappa}\right)^{ab}\Bigg(
        \W_\mu^b + \g \w_\mu^b + g^\prime\PL_{\mu\nu} \b_\nu \delta^{b3}
        - {2 \sqrt{\lambda}\over m} \left(\varD_\mu {\eta\over
	R}\right)^b 
        -  {2 \lambda\over m^2} \varepsilon^{bcd} {\eta^c\over R}
        \left(\varD_\mu {\eta\over R}\right)^d \Bigg) \, , 
\eea
where 
\be
\varD_\mu^{ab} \eta^b = \p_\mu \eta^a - \varepsilon^{abc} \W_\mu^c
\eta^b \ . 
\ee

The basic idea for this choice of parametrization of the quantum
fluctuations is the following. If we shift the fields only linearly,
the correspondence between zero modes of the differential operator and
fluctuations corresponding to gauge transformations is only true at
leading order.  However, if we use the parametrization given above
this correspondence is true at higher orders in the fluctuations as
well.

We collect all the fluctuations in a vector
\be \label{fluct_vec}
y = \left( \begin{array}{c} 
                f \\
                \eta^a \\
                \tq_\mu^A
\end{array}\right)   \ ,
\ee
where $\tq_\mu^A \doteq \left( \begin{array}{c} \w_\mu^a \\ \b_\mu
\end{array} \right)$ describes the fluctuations of the gauge fields.
Here and in the following, lowercase Latin indices ($a,b$) run from
$1$ to $3$, whereas uppercase Latin indices ($A,B$) run from $1$ to
$4$. The differential operator $\tildeD$ which is acting on the space of
fluctuations from Eq.~(\ref{fluct_vec}) can be represented by
a $3\times 3$-matrix. Before we write down this matrix it is useful to
make some additional transformations of the differential operator. 

As noted above, the differential operator $\tildeD$ has zero modes due
to gauge invariance. In the basis $f, \eta^a, \tq_\mu^A$ they can be
written in the form 
\begin{equation} \label{zeromodes}
            \left( \begin{array}{c}
                0 \\
                M_W R \delta^{aB} \\
                \tvarD_\mu^{AB} 
              \end{array} \right) \alpha^B
            \equiv P \alpha \, , 
\end{equation}
where $\alpha^B$ are four arbitrary scalar functions. The covariant
derivative $\tvarD_\mu^{AB}$ is defined through
\bea 
\tvarD_\mu^{AB} & = & \delta^{AB} \p_\mu - f^{ABc} \W_\mu^c \, ,
\label{def_tvarD} \\
f^{ABc}         & = & \left\{ \begin{array}{ccl}
                                \varepsilon^{abc} & , &  A=a, B=b \, , 
                                \\
                                0 & , & A=4 \ \mbox{and / or} \ B=4 \, . 
                             \end{array} \right. 
\eea

The generating functional is then given by the
expression in Eq.~(\ref{1loopgf}). Using the fact that zero and non-zero
eigenvectors are orthogonal to each other leads to the identity 
\begin{equation}
  {\ln\det}^\prime \tildeD = \ln\det\left(\tildeD + P P^T +
  \delta_P\right) - \ln\det(P^T P) \ , 
\end{equation}
up to an irrelevant infinite constant.  Again, $\det^\prime \tildeD$
denotes the product of all non-zero eigenvalues. The operator
$\delta_P$ will be defined below. 

In order to remove the dependence of the differential operators
$\tildeD + P P^T + \delta_P$ and $P^T P$ on the phase factor $\phase$
we define  
\bea
\uDfull & \doteq & O_1 \left( \tildeD + P P^T + \delta_P \right) O_1^T 
\, , 
\label{def_uDfull} \\ 
\underline{P^T P} & \doteq & O_2 (P^T P) O_2^T \, , 
\label{def_uPTP} 
\eea
where
\be
O_1 = 
\left( \begin{array}{ccc} 
1 & 0 & 0 \\
0 & \phase^{ab} & 0 \\
0 & 0 & O_2^{AB} 
\end{array} \right) , \quad 
O_2  =  
\left( \begin{array}{cc} 
\phase^{ab} & 0  \\
0 & 1  \\
\end{array} \right) . 
\ee
The transformation matrices $O_1$ and $O_2$ have unit determinant
since $\det\phase=1$, because the matrix~$T$ which appears in the phase
factor is traceless, cf.\ Eq.~(\ref{matrix_T}). 

Therefore, the generating functional at the one-loop level can be
written in the following form:  
\begin{equation} \label{1loopgf_1}
        W_\SM[h,K_{\mu\nu},J_\mu^a] = \intdx \lag_\SM
           + \half {\ln\det}\left(\uDfull \right) - \ln\det
           \underline{P^T P} \, , 
\end{equation}
where the solutions of the equations of motion
(\ref{eomRcomp})--(\ref{eomAcomp}) have to be inserted. It represents
the full one-loop contributions of the bosonic sector of the standard
model.

The explicit expressions for the components of the differential
operator $\uDfull$, which we parametrize by 
\be
\uDfull \doteq  \left( \begin{array}{ccc}
                        d       & \delta        & \delta_\nu  \\
                    \delta^T    & D             & \Delta_\nu  \\
                   \delta_\mu^T & \Delta_\mu^T  & D_{\mu\nu}
                \end{array} \right) , \label{defD} \\
\ee  
can be found in Eqs.~(\ref{firstcomp_defD})--(\ref{lastcomp_defD})
in Appendix~\ref{app:diffop}. The operators $\underline{P P^T},
\underline{P^T P}$, and $\delta_P$ are listed in
Eqs.~(\ref{def_PPT})--(\ref{def_deltaP}).    

We would like to stress an important point here. At the classical
level only physical modes propagate. The classical Goldstone boson
field $U^{cl}$ represents the $SU(2)_L$ gauge degrees of freedom. At
the quantum level, however, the situation is different. Quantum
fluctuations around the classical field $U^{cl}$ denoted by $\eta^a$,
cf.\ Eqs.~(\ref{fluct_U}) and (\ref{def_V}), imply virtual Goldstone
boson modes propagating within loops. Note that these modes are absent
in any gauge-dependent approach based on the unitary gauge.  They are,
however, necessary in order to ensure a decent high-energy behavior
of the theory.
 
In order to diagonalize the differential operator $\uDfull$ we
introduce some additional quantities:   
\bea
\varD_{\mu\nu} & = & D_{\mu\nu} - \delta_\mu^T d^{-1} \delta_\nu
               - \vartheta_\mu^T \Theta^{-1} \vartheta_\nu \, , 
\label{varD} \\
\Theta         & = & D - \delta^T d^{-1} \delta \, , \\
\vartheta_\nu  & = & \Delta_\nu - \delta^T d^{-1} \delta_\nu \ . 
\label{vartheta} 
\eea
Using the identity 
\be \label{trafowithT}
\Trafo^T \left( \uDfull \right) \Trafo =
\mathrm{diag} \left( d, \, \Theta, \, \varD_{\mu\nu} \right) \, ,  
\ee
where 
\be
\Trafo         =  \left( \begin{array}{ccc}
  1 & - d^{-1} \delta  & - d^{-1} \delta_\nu + d^{-1} \delta
                          \Theta^{-1} \vartheta_\nu   \\ 
  0 & 1  & - \Theta^{-1} \vartheta_\nu \\
  0 & 0  & \delta_{\mu\nu}
\end{array} \right) \, , 
\ee
and the fact that the transformation matrix $\Trafo$ has unit
determinant, one obtains the following intermediate result for the
generating functional:   

\be \label{1loopgf_2}
      W_\SM[h,K_{\mu\nu},J_\mu^a] = \intdx \lag_\SM
           + \half \ln\det d + \half \ln\det \Theta 
+ \half \ln\det \varD - \ln\det \underline{P^T P} \ .  
\ee

In a second step we rotate from the fluctuations of the weak
eigenstates of the gauge bosons to the corresponding mass eigenstates: 
\begin{equation} \label{rotation}
  \varD_{\mu\nu} \rightarrow \widetilde\varD_{\mu\nu} = O
  \varD_{\mu\nu} O^T \, , 
\end{equation}
with the orthogonal matrix 
\be \label{trafowithO}
O = \left(\begin{array}{cccc}
1 & 0 & 0 & 0 \\
0 & 1 & 0 & 0 \\
0 & 0 & c & -s \\
0 & 0 & s & c 
\end{array}\right) \, . 
\ee 
After this rotation we get 
\be \label{tildevarD}
 \widetilde\varD_{\mu\nu} = 
\left(\begin{array}{ccc}
 \widetilde D_{\mu\nu}^\W & \widetilde\xi_{\mu\nu}^\Z &
                                          \widetilde\xi_{\mu\nu}^\A \\
 \widetilde\xi_{\mu\nu}^{\Z, T} &
                                          \widetilde D_{\mu\nu}^\Z &
                                          \widetilde\eta_{\mu\nu}^\A \\
 \widetilde\xi_{\mu\nu}^{\A, T} & 
                                \widetilde\eta_{\mu\nu}^{\A, T} &
                                \widetilde D_{\mu\nu}^\A 
\end{array}\right)  \ ,
\ee
where the components are defined by the following equations: 
\bea
 \widetilde D_{\mu\nu}^M  & = & D_{\mu\nu}^M 
- \delta_\mu^{M, T} d^{-1} \delta^M_\nu 
- \vartheta_\mu^{M, T} \Theta^{-1} \vartheta^M_\nu \ , \ M = \W, \Z,
\A \, , 
\label{tildeDmunuW}\\
 \widetilde\xi_{\mu\nu}^M & = & \xi_{\mu\nu}^M 
- \delta_\mu^{M, T} d^{-1} \delta^M_\nu 
- \vartheta_\mu^{M, T} \Theta^{-1} \vartheta^M_\nu \ , \ M = \Z, \A \, 
, \\
 \widetilde\eta_{\mu\nu}^\A & = & \eta_{\mu\nu}^\A
- \delta_\mu^{\A, T} d^{-1} \delta^\A_\nu 
- \vartheta_\mu^{\A, T} \Theta^{-1} \vartheta^\A_\nu \ . 
\label{tildeetamunuA} 
\eea
Similar relations hold for the transposed components
$\widetilde\xi_{\mu\nu}^{\Z, T}, \widetilde\xi_{\mu\nu}^{\A, T},$ and
$\widetilde\eta_{\mu\nu}^{\A, T}$. Furthermore, we have introduced the
quantities 
\be
\vartheta^M_\mu =  \Delta_\mu^M - \delta^T d^{-1} \delta_\mu^M,
\qquad M = \W, \Z, \A \ . \label{def_varthetaM} 
\ee
Note that the index $\W$ which appears in Eqs.~(\ref{tildevarD}),
(\ref{tildeDmunuW}), and (\ref{def_varthetaM}) refers to the two
components $\w_\mu^{1,2}$ of the fluctuations which correspond to the
original fields $W_\mu^{1,2}$. We will use Greek letters $\alpha,
\beta = 1,2$ to label these two components.  
The expressions for the differential operators which appear in
Eqs.~(\ref{tildeDmunuW})--(\ref{tildeetamunuA}) can be found in
Appendix~\ref{app:diffop}. The operators $\delta_\mu$ and $\Delta_\mu$
are listed in Eqs.~(\ref{delta_nu_W})--(\ref{Delta_mu_A_T}). The first
terms on the right-hand side of
Eqs.~(\ref{tildeDmunuW})--(\ref{tildeetamunuA}) are given by
Eqs.~(\ref{firstcomp_tildevarD})--(\ref{lastcomp_tildevarD}).

Finally, we can diagonalize the differential operator
$\tvarD_{\mu\nu}$ from Eq.~(\ref{tildevarD}) in an analogous way to
the diagonalization of the operator $\uDfull$ in
Eq.~(\ref{trafowithT}).  The corresponding transformation matrix has
again unit determinant.  This is, of course, also true for the
orthogonal matrix $O$, Eq.~(\ref{trafowithO}).  Therefore we can write
the generating functional in the following form:
\bea 
W_\SM[h,K_{\mu\nu},J_\mu^a] & = & \intdx
\lag_\SM  + \half \ln\det d + \half \ln\det \Theta 
+ \half \ln\det \varD^\W + \half \ln\det \varD^\Z + \half
\ln\det \varD^\A - \ln\det \underline{P^T P} \ . \nonumber \\
& & \label{1loopgf_3} 
\eea
The operators which appear in Eq.~(\ref{1loopgf_3}) are defined
through the relations 
\begin{eqnarray}
\varD_{\mu\nu}^\W & \equiv & \widetilde D_{\mu\nu}^\W 
= D_{\mu\nu}^\W - \delta_\mu^{\W, T} d^{-1} 
\delta_\nu^\W - \vartheta_\mu^{\W,T} \Theta^{-1} \vartheta_\nu^\W \, , 
\label{def_varDW} \\
\varD_{\mu\nu}^\Z &=& \widetilde D_{\mu\nu}^\Z 
 - \widetilde\xi_{\mu\rho}^{\Z, T} \left({\varD^\W}\right)^{-1}_{\rho\sigma}
  \widetilde\xi_{\sigma\nu}^\Z \, , \\
\varD^\A_{\mu\nu} &=& \widetilde D_{\mu\nu}^\A 
 - \widetilde\xi_{\mu\rho}^{\A, T} \left({\varD^\W}\right)^{-1}_{\rho\sigma}
  \widetilde\xi_{\sigma\nu}^\A  
- N_{\mu\rho}^{\A, T} \left({\varD^\Z}\right)^{-1}_{\rho\sigma}
  N_{\sigma\nu}^\A \, , \\
N_{\mu\nu}^\A &= & \widetilde \eta_{\mu\nu}^\A  
   - \widetilde \xi_{\mu\rho}^{\Z, T}
   \left(\varD^\W\right)^{-1}_{\rho\sigma}\widetilde\xi^\A_{\sigma\nu}
   \ . 
\end{eqnarray}

Equation~(\ref{1loopgf_1}) and the equivalent forms in
Eqs.~(\ref{1loopgf_2}) and (\ref{1loopgf_3}) represent our result for
the generating functional $W_{\SM}[h,K_{\mu\nu}, J_\mu^a]$ for the
gauge-invariant Green's functions for the bosonic sector of the
standard model. These formulas encode the full tree-level and one-loop
effects of the theory. If one expands the generating functional up to
a given order in powers of the external sources one can extract any
$n$-point Green's functions for the gauge-invariant operators
$\Bphi^\dagger \Bphi, B_{\mu\nu},$ and $\u V_\mu^a$. Since the
equations of motion~(\ref{eomRcomp})--(\ref{eomUcompZ}) only involve
gauge-invariant fields and because the differential operators which
enter the generating functional only contain gauge-invariant
quantities, cf.\ the explicit expressions in
Appendix~\ref{app:diffop}, the evaluation of the generating functional
and the final result will be manifestly gauge-invariant. In
Sec.~\ref{sec:greensfunctions} we will calculate the two-point
functions of these gauge-invariant operators. Before we come to this,
we discuss in the next section the renormalization of the theory in
the presence of the external sources.


\section{Renormalization}
\label{sec:renorm}

In order to render the generating functional
$W_\SM[h,K_{\mu\nu},J_\mu^a]$ of the standard model in
Eq.~(\ref{1loopgf_1}), and the equivalent forms in
Eqs.~(\ref{1loopgf_2}) and (\ref{1loopgf_3}), finite, one has to
renormalize the bare constants $m^2, \lambda, \g, \gp,$ the scalar
field $\Bphi$ and the sources before the regulator can be removed.
There is no wave-function renormalization for the gauge fields
$W_\mu^a$ and $B_\mu$ on account of gauge invariance, cf.\ our
definition of the covariant derivative in Eq.~(\ref{cov_deriv}). The
ultraviolet divergences are related to the poles of the
$d$-dimensional determinant, which appear in the generating functional
for $d=0,2,4, \ldots$. In general, for a differential operator $\bar
D$ of the form
\begin{equation} \label{Dform}
  \bD = -\bD_\mu \bD_\mu + \bsigma \ , \qquad \bD_\mu
  = \partial_\mu + \bGamma_\mu \ ,
\end{equation}
the pole term of the determinant at $d=4$ is given by
\begin{equation} \label{Det_pole}
        \half \ln\det \bD = {1 \over d-4} {1\over 16\pi^2}
        \intdx \, \tr \left( {1\over12} \bGamma_{\mu\nu} \bGamma_{\mu\nu}
                        + \half \bsigma^2 \right) + \order(1) \ ,
\end{equation}
with
\begin{equation}
        \bGamma_{\mu\nu} = \left [ \bD_\mu , \bD_\nu \right ] \ .
\end{equation}
This identity can readily be derived~\cite{LSM} using the heat kernel
method~\cite{heatkernel}.  The symbol ``tr'' in Eq.~(\ref{Det_pole})
denotes the trace over internal and Lorentz indices which will be
suppressed in the following. We cannot apply, however,
formula~(\ref{Det_pole}) in our case. This is due to the fact that the
differential operator\footnote{For the calculation of the counterterms
  it is convenient to introduce a real representation for the Higgs
  field. See Appendix~\ref{app:renorm_prescriptions} for details.
  }~$\Dfull$, which appears in the one-loop expression for the
generating functional, is not an ordinary local differential operator
of the form given in Eq.~(\ref{Dform}), but a nonlocal functional of
the fields due to the presence of derivatives of the massless
propagator $\G_0(x-y) = \langle x | 1 /(- \Box) | y \rangle$ in
some of the components.  These massless propagators originate from the
phase factor $\phase$, cf.\ Eq.~(\ref{phase_matrix}).  In order to
calculate the counterterms, we therefore split the differential
operator in a local, $\Dloc$, and a ``nonlocal,'' $\dDnl$, part
\be \label{loc_nonloc}
\Dfull = \Dloc + \dDnl \ , 
\ee
where $\dDnl$ contains all the terms stemming from the phase factor. 
The explicit expressions for these operators can be found in
Appendix~\ref{app:renorm_prescriptions}, cf.\ Eqs.~(\ref{Dloc}) and
(\ref{dDnl}). Using this decomposition we can write 
\be
\half \ln\det (\Dfull) =  
\half \ln\det \Dloc
+ \half \Tr \left[ \Dlocinvers \dDnl \right]
- {1 \over 4} \Tr \left[ \left( \Dlocinvers \dDnl \right)^2 \right]
+{1 \over 6} \Tr \left[ \left( \Dlocinvers \dDnl \right)^3 \right]
+ \cdots \, . \label{lndetDnonlocal} 
\ee
The first term on the right-hand side of Eq.~(\ref{lndetDnonlocal})
can now be treated in the usual way using Eq~(\ref{Det_pole}), whereas
we can use a short distance expansion to extract the divergent and
local contributions from the other terms.  Techniques for performing
such a short distance expansion have been discussed in detail in
Ref.~\cite{LSM}. Here we present only the general procedure.

We write the second term on the right-hand side of
Eq.~(\ref{lndetDnonlocal}) in the form 
\be \label{HKM_tadpole}
\half \Tr \left[ \Dlocinvers \dDnl \right] = 
\half \int \d^dx \d^dy \, \tr \left( \langle x | \Dlocinvers | y \rangle 
\langle y | \dDnl | x \rangle \right) \ . 
\ee
Due to the presence of one propagator $\Dlocinvers$ this term can be  
viewed as a tadpole graph. 

The ultraviolet poles in Eq.~(\ref{HKM_tadpole}) are connected to the
short distance properties of the operator $\Dloc$, which in turn are
governed by the Laplacian $\Box$, since $\Dloc$ is of the
form given in Eq.~(\ref{Dform}).  Observing that in $d$-dimensional
Euclidean space one has the identity
\begin{equation}
\langle x | e^{ \lambda \Box} | y \rangle = 
             (4 \pi \lambda)^{-d/2} e^{- z^2 / 4 \lambda}  \ , 
\end{equation}
where $z = x-y$, we define the heat kernel $H(x|\lambda|y)$ through
the equation 
\begin{equation} \label{HK}
\langle x | e^{- \lambda \Dloc} | y \rangle \doteq  (4 \pi \lambda)^{-d/2}  
\ e^{-z^2 / 4 \lambda} \ \ H(x|\lambda|y) \ .
\end{equation}
Using Eq.~(\ref{HK}) we can then write 
\bea
\langle x | \Dlocinvers | y \rangle & = & \int_0^{\infty} {\d \lambda \over 
(4 \pi \lambda)^{d/2} } \ e^{- z^2 / 4 \lambda} \ \ H(x|\lambda|y) \\
& = & \sum_{n=0}^{\infty} \int_0^{\infty} {\d \lambda \over 
(4 \pi \lambda)^{d/2} } \ e^{- z^2 / 4 \lambda} \lambda^n  H_n(x|y) 
\ . 
\label{heatcoeff} 
\eea
In the second line we have expanded the heat kernel $H(x|\lambda|y)$
in terms of the heat coefficients $H_n(x|y)$.  These heat coefficients
can in turn be expanded around the point $x$ in order to obtain local
counterterms at the end: 
\be \label{taylor_z} 
H_n(x|y) = H_n(x|x) - z_\mu \left. (\p_\mu H_n) \right|_{x=y} +
\half z_\mu z_\nu \left. (\p_\mu \p_\nu H_n) \right|_{x=y} + \cdots \, 
 . 
\ee
One now inserts the resulting expression for $\langle x | \Dlocinvers |
y \rangle$ into Eq.~(\ref{HKM_tadpole}). Moreover, one can use the
following representations for the massless propagator $\G_0$ and
derivatives thereof  
\bea
\G_0(z) & = & \int_0^{\infty} {\d
  \rho \over  (4 \pi \rho)^{d/2} } \ e^{- z^2 / 4 \rho} \, , \\
\p_\mu \G_0(z) & = & \int_0^{\infty} {\d
  \rho \over  (4 \pi \rho)^{d/2} } \ \left( - {1\over 2} z_\mu {1\over
  \rho} \right) e^{- z^2 / 4 \rho} \, , \\ 
\p_\mu \p_\nu \G_0(z) & = & \int_0^{\infty} {\d
  \rho \over  (4 \pi \rho)^{d/2} } \ \left( - {1\over 2} \delta_{\mu\nu}
{1\over \rho} + {1\over 4} z_\mu z_\nu {1\over \rho^2} \right)
e^{- z^2 / 4 \rho} \, , 
\eea
which appear in $\langle y | \dDnl | x \rangle$.  If one now performs
the integration over $d^dz$ and the parameters $\rho$ and $\lambda$,
one observes that the ultraviolet poles manifest themselves as
divergences at the lower end of the integration over $\lambda$.  Power
counting shows that higher order heat coefficients $H_n$ lead to a
less singular behavior for short distances since they are accompanied
by additional powers of $\lambda$ in Eq.~(\ref{heatcoeff}).  The same
is true for higher derivative terms in the expansion of $H_n$ in
Eq.~(\ref{taylor_z}), due to the presence of additional powers of
$z_\mu$. Therefore only a finite number of terms in the expansions in
Eqs.~(\ref{heatcoeff}) and (\ref{taylor_z}) lead to ultraviolet
divergent contributions. At the end, we need only the leading order
term $H_0(x|x)$ and the next-to-leading order terms $\left.
(\p_\mu\p_\nu H_0) \right|_{x=y}$ and $H_1(x|x)$ to extract the
counterterms.  The explicit expressions for these quantities can be
inferred from the results (cf.\ Ref.~\cite{LSM} for the derivation):
\bea
H_0(x|x) & = &  1 , \quad  
H_1(x|x) = - \bsigma \, , \nonumber \\
\left.(\bD_\mu H_0)\right|_{x=y} & = & 0 , \quad  
\left.\left( \bD_\mu \bD_\nu H_0 \right)\right|_{x=y} =     
\half \bGamma_{\mu \nu} \, . 
\eea

\noindent
The third term in Eq.~(\ref{lndetDnonlocal}) can be written in the
  form  
\be
- {1\over 4} \Tr \left[ (\Dlocinvers \dDnl)^2 \right] = 
- {1\over 4} \int \d^dx \d^dy \d^d z \d^d u \, \tr \left(  
\langle x | \Dlocinvers | y \rangle 
\langle y | \dDnl | z \rangle 
\langle z | \Dlocinvers | u \rangle 
\langle u | \dDnl | x \rangle \right) \, , \label{HKM_twopoint} 
\ee
which can be interpreted as a two-point function with two
propagators $\Dlocinvers$. Similar arguments as used above then lead to the
observation that we need only the leading term in the short distance
behavior of $\Dloc$, which amounts to setting 
\be \label{leadingorder} 
\langle x | \Dlocinvers | y \rangle \to 
\langle x | {1\over -\Box} | y \rangle \ , 
\ee
in Eq.~(\ref{HKM_twopoint}). Again we have suppressed all
internal and Lorentz indices in Eq.~(\ref{leadingorder}).  Finally,
all the subsequent terms in Eq.~(\ref{lndetDnonlocal}), which contain
three and more propagators $\Dlocinvers$, lead to ultraviolet finite
integrals.

The ultraviolet contributions from the measure of the path integral to
the generating functional, i.e.\ the term $- \ln\det \bP^T \bP$, can be
treated in the usual way, using Eq.~(\ref{Det_pole}). The explicit
result for the operator $\bP^T \bP$ can be found in Eq.~(\ref{bPTbP}).

This procedure leads to the counterterm Lagrangian from which one can
read off the renormalization prescriptions which will remove the poles
in the generating functional in Eq.~(\ref{1loopgf_1}). The full list
of renormalization prescriptions for all the fields, the mass
parameter $m^2$, the coupling constants, and the sources are listed in
Appendix~\ref{app:renorm_prescriptions} in
Eqs.~(\ref{W_mu_ren})--(\ref{gprime_ren}), and
(\ref{firstsource_ren})--(\ref{lastrenormprescription}).  In the next
section we will only need the renormalization prescriptions for the
fields, the mass and the coupling constants which are given in
Eqs.~(\ref{W_mu_ren})--(\ref{gprime_ren}).


\section{Physical input parameters from gauge-in\-variant Green's
  functions}
\label{sec:greensfunctions}

In this section we relate the bare parameters of the theory to
physical quantities. As physical input parameters we choose the masses
of the Higgs and the $W$ and $Z$ bosons, and the electric charge
(on-shell scheme). The physical mass of the Higgs boson, which we
denote by $\MHps$, is determined by the pole position of the two-point
function
\be \label{twopoint_Higgs}
\langle 0 | T (\Bphi^\dagger \Bphi)(x) (\Bphi^\dagger \Bphi)(y)| 0
\rangle \ . 
\ee
The physical masses of the $W$-boson, $\MWps$, and the $Z$-boson,
$\MZps$, are defined by the pole positions of the two-point function
\be \label{twopoint_V}
\langle 0 | T (\u V_\mu^a)(x) (\u V_\nu^b)(y)| 0 \rangle \ . 
\ee
The electric charge is determined by the three-point function 
\be \label{threepoint_BVV} 
\langle 0 | T \B_{\mu\nu}(x) (\u V_\rho^+)(y) (\u V_\sigma^-)(z) |
 0 \rangle \ . 
\ee
Alternatively, one can define a renormalized electric charge as the
residue at the photon pole of the two-point function  
\be \label{twopoint_Bmunu} 
\langle 0 | T B_{\mu\nu}(x) B_{\rho\sigma}(y)| 0 \rangle \ . 
\ee
We will denote the corresponding coupling constant by $\ers$. At this
point several comments are in order. In the usual approach to gauge
theories dealing with gauge-dependent Green's functions, two- and
three-point functions are related by Ward identities. These kind of
Ward identities follow, however, from the fact that these Green's
functions are gauge-dependent. In contrast, there are no Ward
identities of this type between our gauge-invariant Green's functions.
Nevertheless, the absence of Ward identities does not imply any lack
of knowledge. All information that Ward identities encode in any
gauge-dependent framework is manifest in our approach. Therefore, one
may expect the coupling constant which can be extracted from the
three-point function in Eq.~(\ref{threepoint_BVV}) and from the
two-point function in Eq.~(\ref{twopoint_Bmunu}) to be the same, if
they are evaluated at the same scale.  Below, we will extract the
electric charge $\ers$ from the residue of the two-point function in
Eq.~(\ref{twopoint_Bmunu}).  We have not explicitly checked whether
the definition through the three-point function in
Eq.~(\ref{threepoint_BVV}) leads to the same result.  Note that the
residue of the two-point function of the field strength $B_{\mu\nu}$
in Eq.~(\ref{twopoint_Bmunu}) differs from unity and that there is no
freedom to adjust the residue by a renormalization factor.  This can
be traced back to our normalization of the gauge field $B_\mu$ in the
covariant derivative, cf.\ Eq.~(\ref{cov_deriv}).

Before we begin with the evaluation of the physical input parameters
let us discuss an example for relations that usually are derived with
the help of Ward identities, but which are manifest in our approach.
Consider the covariant derivative $D_\mu$ as defined in
Eq.~(\ref{cov_deriv}).  Gauge invariance ensures that the covariant
structure of $D_\mu$ is not destroyed by coun\-ter\-terms. In fact all
counterterms are gauge-invariant in our approach. Therefore the fields
$W_\mu^a$ and $B_\mu$ are not renormalized, cf.\ Eqs.~(\ref{W_mu_ren})
and (\ref{B_mu_ren}).  Factoring out the gauge coupling constants $g$
and $\gp$ from the gauge fields, which we then denote by $\hat
W_\mu^a$ and $\hat B_\mu$, the covariant derivative reads
\be
D_\mu \Bphi = \left( \p_\mu - i {\tau^a\over 2} g \hat W_\mu^a 
                          - i {1\over 2} \gp \hat B_\mu \right) \Bphi
\, . 
\ee
Thus, the wave function renormalization of the field $\hat W_\mu^a$
($\hat B_\mu$) must be the inverse of the renormalization for $g$
($\gp$). This follows automatically from gauge invariance. 

For the determination of the two-point functions in
Eqs.~(\ref{twopoint_Higgs}), (\ref{twopoint_V}), and
(\ref{twopoint_Bmunu}) we need the generating functional
$W_\SM[h,K_{\mu\nu},J_\mu^a]$ up to second order in the external
sources. Using a saddle-point approximation for the path integral, the
generating functional at tree level is given by the action,
evaluated at the solutions of the classical equations of motion.
Inserting the solutions of the equation of motion from 
Eqs.~(\ref{solution_R})--(\ref{solution_AT}) into the classical
action we get the following result for the generating functional at
tree level, expanded up to second order in powers of the external
sources: 
\bea 
W_\SM[h,K_{\mu\nu},J_\mu^a]^{\mathrm{tree}} 
& = &\int d^dx \, \Bigg\{ - {m^2\over 2\lambda} h(x) \Bigg\} 
+ \int d^dx d^dy \, \Bigg\{ 
- {m^2\over 2\lambda} h_x \G_H(x-y) h_y 
- {8m^2\over\lambda} M_W^2 \left( J_{\mu,x}^{+,T} \G_W(x-y)
J_{\mu,y}^{-,T}  \right)
\nonumber  \\ 
& &\mbox{}- {s^2 c^2\over 2 e^2} \left({e^2\over c^2} (\p_\nu \widehat
K_{\nu\mu}) - 4 M_Z^2 J_\mu^{\Z,T} \right)_x \G_Z(x-y)
\left({e^2\over c^2} (\p_\rho \widehat K_{\rho\mu}) - 4 M_Z^2
J_\mu^{\Z,T} \right)_y   
\nonumber \\ 
&&\mbox{}- \half e^2 (\p_\nu \widehat K_{\nu\mu})_x \G_A(x-y)
(\p_\rho \widehat K_{\rho\mu})_y + \mbox{contact terms} 
\Bigg\} \ . \label{W_tree_sources}
\eea
We have indicated the space-time arguments of the sources by the
subscripts $x$ and $y$. 

The contact terms in Eq.~(\ref{W_tree_sources}) do not contribute to
the pole positions of the two-point functions. Note that only the
transversal modes are propagating at this order of the expansion. From the
pole positions of the propagators we can read off the masses of the
particles at tree level:
\be
M_H^2 = 2m^2; \ M_W^2; \ M_Z^2; \ M_\gamma^2 = 0 \ . 
\ee
The two-point function of the field strength $B_{\mu\nu}$ in
Eq.~(\ref{twopoint_Bmunu}) is obtained from the generating functional
in Eq.~(\ref{W_tree_sources}), if we switch off all the sources, except
for $K_{\mu\nu}$. It contains poles at $p^2 = 0$ and at $p^2 = -
M_Z^2$ (note that we are working in Euclidean space-time), because of
the presence of the propagators $\G_A$ and $\G_Z$ in
Eq.~(\ref{W_tree_sources}). This is due to the fact that the field
$B_\mu$ is a linear combination of the photon and the $Z$-boson field,
cf.\ Eq.~(\ref{bfieldsl}). From the residue at the pole position of
the photon we get the electric charge
\be \label{e_tree}
\ers = e^2 \ . 
\ee

At the one-loop level the generating functional is given by the
expression in Eq.~(\ref{1loopgf_3}), where the diagonalization of
the differential operator for the quantum fluctuations has been
carried out completely. Expanding the determinants in powers of the
external sources leads after a lengthy calculation to the following
result for the generating functional:
\bea 
W_\SM[h,K_{\mu\nu},J_\mu^a] & = & \int \ddp \Bigg\{
- {m^2\over 2\lambda} h(p)   
- \left( {m^2 \over 2 \lambda} \right) h(p) \G_H(p) \left[ 1 +
\Sigma_{\Bphi^\dagger\Bphi}(p^2) \G_H(p) \right] h(-p) \nonumber  \\ 
& &\mbox{}- \left( {8m^2\over\lambda} M_W^2 \right) J_\mu^{+,T}(p) \G_W(p)
\left[ 1 + \Sigma_\uW^T(p^2) \G_W(p) \right] J_\mu^{-,T}(-p) \nonumber \\ 
& &\mbox{}- \half \left( \widetilde J_\mu^{\Z,T}(p), \widetilde
J_\mu^{\A,T}(p) \right)
\G_{Z\gamma}(p) \left[{\mathbf{1}} +
{\mathbf{\Sigma}}_{\Z\gamma}^T(p^2) \G_{Z\gamma}(p) \right]  
\left(\begin{array}{c}
\widetilde J_\mu^{\Z,T}(-p) \\
\widetilde J_\mu^{\A,T}(-p)
\end{array} \right) 
\nonumber \\
& &\mbox{}+ 16 J_\mu^{+,L}(p) \Sigma_\uW^L(p^2) J_\mu^{-,L}(-p)
+ 4 J_\mu^{\Z,L}(p) \Sigma_\Z^L(p^2) J_\mu^{\Z,L}(-p) 
+ \mbox{contact terms} 
\Bigg\} \, , 
\label{W_1loop_sources}
\eea
where we have introduced the abbreviations 
\bea
\widetilde J_\mu^{\Z,T}(p) & = & 
{s c\over e} \left(- 4 M_Z^2 J_\mu^{\Z,T}(p) +
{e^2\over c^2} (\p_\rho \widehat K_{\rho\mu}(p)) \right) \, , \\
\widetilde J_\mu^{\A,T}(p) & = & e \, (\p_\rho \widehat
K_{\rho\mu}(p)) \, , \\
\G_{Z\gamma}(p) & = & 
\left( \begin{array}{cc}
\G_Z(p) & 0 \\
0 & \G_A(p)  
\end{array} \right) \, , \\
{\mathbf{\Sigma}}_{\Z\gamma}^T(p^2) & = & 
\left( \begin{array}{cc}
\Sigma_\Z^T(p^2) & \Sigma_{\Z \A}^T(p^2) \\
\Sigma_{\A \Z}^T(p^2) & \Sigma_\A^T(p^2)  
\end{array} \right) \, , \label{Sigma_Zg} \\
\mathbf{1} & = & 
\left( \begin{array}{cc}
1 & 0 \\
0 & 1  
\end{array} \right) . 
\eea
The propagators $\G$ in Eq.~(\ref{W_1loop_sources}) are defined in
Eq.~(\ref{treelevel_prop}).  The explicit results for the
self-energies $\Sigma$ can be found in
Appendix~\ref{app:selfenergies}. We note that these self-energies are
part of our gauge-invariant Green's functions and should not be
confused with the self-energies of the Higgs boson and the gauge
fields in the usual approach.

Since our calculation preserves gauge invariance, the self-energy
$\Sigma_\A^T$ from Eq.~(\ref{def_SigmaT_A}) has the property
$\Sigma_\A^T(0) = 0$ which guarantees that the photon remains
massless.  From the result in Eq.~(\ref{def_SigmaT_ZA}) follows that
$\Sigma_{\Z\A}^T(0) = 0$.  Therefore the self-energy mixing-matrix
${\mathbf{\Sigma}}_{\Z\gamma}^T(p^2)$ in Eq.~(\ref{Sigma_Zg}) is
diagonal at $p^2=0$.  Furthermore, gauge invariance implies that the
equations of motion and the differential operators only contain the
transverse component $\A_\mu^T$. Therefore only the self-energies
$\Sigma_{\Z\A}^T$ and $\Sigma_\A^T$ appear in the generating
functional in Eq.~(\ref{W_1loop_sources}). There are no quantities
$\Sigma_{\Z\A}^L$ and $\Sigma_\A^L$. All these properties of the
self-energies follow directly from gauge invariance. There is no need
to impose them by any kind of renormalization conditions.  We note
that these properties for the self-energies also hold for the
corresponding self-energies of the gauge fields in the background
field approach to the standard model~\cite{BGFM_SM}.  The latter
Green's functions still depend, however, on the gauge fixing parameter
$\xi_Q$ for the quantum fluctuations.

As a check on our calculation we have verified that in the limit $\gp
\to 0$ the self-energies $\Sigma_\Z$ and $\Sigma_\uW$ coincide.  This
statement is true for the transversal self-energies, $\Sigma^T$,
Eqs.~(\ref{def_SigmaT_W}) and (\ref{def_SigmaT_Z}), as well as the
longitudinal self-energies $\Sigma^L$, Eqs.~(\ref{def_SigmaL_W}) and
(\ref{def_SigmaL_Z}).

From the expression for the generating
functional in Eq.~(\ref{W_1loop_sources}) we define the full
propagators:   
\bea
\G^{\mathrm{full}}(p^2) & = & \G(p^2) \left( 1 + \Sigma(p^2) \G(p^2)
\right) \label{G_full} \\
& = & {1\over p^2 + M^2 - \Sigma(p^2)} \ . \label{G_full_resummed} 
\eea
The second line follows after Dyson resummation. 
In Eq.~(\ref{G_full_resummed}) we denoted the bare
mass by $M^2$. For the gauge bosons we consider only the transverse
components. The definition in Eq.~(\ref{G_full}) can be applied
to the mixing propagator $\G_{Z\gamma}^{\mathrm{full}}$, as
well.  

Using the explicit results from Appendix~\ref{app:selfenergies} one
observes that all self-energies behave for large momenta as
\be
\Sigma(p^2) \sim p^2 \ln ( p^2 / \mu^2)
\quad \mbox{for} \ p^2 \to \infty \ . 
\ee
Note that there are individual contributions to the self-energies
which grow like $p^4$. They cancel each other, however, in the large
$p^2$ limit. Therefore, the full propagators in Eq.~(\ref{G_full}) or
(\ref{G_full_resummed}) have the proper high energy behavior
proportional to $1 / (p^2 \ln ( p^2 / \mu^2))$. This can be traced
back to the fact that the Goldstone boson modes are present in the
calculation of the generating functional at one-loop level. The
propagators in Eq.~(\ref{G_full}) or (\ref{G_full_resummed}) are not
identical to the propagators in the unitary gauge in the usually
employed formalism.

We note that there are terms proportional to $1/p^2$ in
$\Sigma_\uW^{T,L}$ and $\Sigma_\Z^{T,L}$, cf.\
Eqs.~(\ref{def_SigmaT_W})--(\ref{def_SigmaL_Z}).  The limit $p^2 \to
0$ in these self-energies is, however, well defined.

We define the physical masses of the particles through the pole
position of the two-point functions in Eqs.~(\ref{twopoint_Higgs}) and
(\ref{twopoint_V}). These poles then appear in the full, resummed
propagators as defined in Eq.~(\ref{G_full_resummed}).  The masses of
the $Z$-boson and the photon can be identified with the eigenvalues of
the inverse full mixing-propagator, i.e.\ the zeros of the determinant
of $(\G_{Z\gamma}^{\mathrm{full}})^{-1}$. At the one-loop level we get
the relations
\bea 
M^2_{H,\mathrm{pole}} & \doteq & 2m^2 - \mathrm{Re} \left[
\Sigma_{\Bphi^\dagger\Bphi}(p^2 = - M^2_{H,\mathrm{pole}}) \right]
\label{def_MH_pole}  \\ 
& \approx & 2m^2 - \mathrm{Re} \left[ \Sigma_{\Bphi^\dagger\Bphi}(p^2
= - 2m^2) \right] \, , \label{MH_pole} \\
M^2_{W,\mathrm{pole}} & \doteq & M_W^2 - \mathrm{Re} \left[
\Sigma_\uW^T(p^2 = - M^2_{W,\mathrm{pole}}) \right] \label{def_MW_pole} \\
& \approx & M_W^2 - \mathrm{Re} \left[ \Sigma_\uW^T(p^2 = -
M_W^2) \right] \, , \label{MW_pole} \\ 
M^2_{Z,\mathrm{pole}} & \doteq & M_Z^2 - \mathrm{Re} \left[
\Sigma_\Z^T(p^2 = - M^2_{Z,\mathrm{pole}}) \right] \label{def_MZ_pole}  \\ 
& \approx & M_Z^2 - \mathrm{Re} \left[ \Sigma_\Z^T(p^2 = -
M_Z^2) \right] \, . \label{MZ_pole}  
\eea
The approximations are valid at the one-loop level. Note that we are
working in Euclidean space-time. Furthermore, only 
bare quantities enter on the right-hand side of the
Eqs.~(\ref{MH_pole}), (\ref{MW_pole}), and (\ref{MZ_pole}). The
photon remains massless due to the relation $\Sigma_\A^T(0) = 0$.  At
the one-loop level only the diagonal elements of the
self-energy~${\mathbf{\Sigma}}_{\Z\gamma}^T$ from Eq.~(\ref{Sigma_Zg})
enter the definitions for the physical masses of $Z$-boson and the
photon. The explicit expressions for the physical masses can be
inferred from the results for the self-energies given in
Appendix~\ref{app:selfenergies}. We do not list them here because they
are too lengthy.  

We define the electric charge $\ers$ as the residue at the photon pole
of the two-point function $\langle 0 | T B_{\mu\nu}(x)
B_{\rho\sigma}(y) | 0 \rangle$. As discussed before, this two-point
function has poles at $p^2 = 0$ and at $p^2 = - M_Z^2$. Due to the
fact that $\Sigma_{\Z\A}^T(0)=0$, the residue at the photon pole is  
given by the expression 
\be \label{def_e_res}
\ers \doteq {e^2 \over 1 - \left. {\p \over \p p^2} \Sigma_\A^T(p^2)
\right|_{p^2 = 0} } \approx e^2 \left( 1 + \left. {\p \over \p p^2}
\Sigma_\A^T(p^2) \right|_{p^2 = 0} \right) \ .
\ee
The approximation used is valid at the one-loop level. From the
expression for $\Sigma_\A^T$ in Eq.~(\ref{def_SigmaT_A}) we get the
following relation between the physical coupling constant $\ers$ and
the bare coupling constant $e^2$:  
\bea
\ers & = & e^2 \left(1 + e^2 \delta e^2 \right) \, ,
\label{e_res_physical} \\  
\delta e^2 & = & -14  \left[ \polem + {1\over 32 \pi^2} \ln \left(
{M_W^2\over 2m^2} \right) \right] - {19\over 3} {1\over 16 \pi^2} \, , 
\eea
where 
\be
\polem   \doteq {\mu^{d-4} \over 16\pi^2} \left( {1 \over d-4} -
{1\over2} (\ln(4\pi) +    \Gamma'(1) + 1) \right) 
+ {1\over 32\pi^2} \ln \left( {2m^2 \over \mu^2} \right) \ . 
\ee
Only bare quantities appear on the right-hand side of
Eq.~(\ref{e_res_physical}). We note that the result for 
$\ers$ agrees with the usual definition of the electric charge in the
Thompson limit~\cite{delta_e_Thompson} in the absence of fermion
contributions. 

The expressions for the physical masses, Eqs.~(\ref{MH_pole}),
(\ref{MW_pole}), (\ref{MZ_pole}), and the electric charge $\ers$,
Eq.~(\ref{e_res_physical}), are finite if one inserts the
renormalization prescriptions from
Eqs.~(\ref{W_mu_ren})--(\ref{gprime_ren}) for the bare quantities. The
cancellation of the pole terms served as an important test of our
calculation.


\section{Summary and Discussion}
\label{sec:summary}

In this article we have presented a manifestly gauge-invariant
approach to the bosonic sector of the standard model. Its essential
feature is to consider gauge-invariant Green's functions. Hence, the
generating functional involves external sources that couple to
gauge-invariant operators only. In order to obtain the same $S$-matrix
elements as in the usual gauge-dependent approach, we chose sources
that emit one-particle states of the Higgs boson, the $W$- and the
$Z$-boson and the photon. In addition to that, however, the off-shell
behavior of our Green's functions is completely free of any
gauge-artifacts. This property makes our approach particularly
suitable for situations where one is interested to gain information
from off-shell quantities or where one is forced to deal with them,
like, for instance, the analysis of finite width effects or the
parametrization of new physics in terms of the oblique parameters
$S,T$ and $U$. 

As gauge-invariant operators we chose the scalar density
$\Bphi^\dagger \Bphi$, the Abelian field strength $B_{\mu\nu}$ and the
quantities $\phase^{ab} V_\mu^b$, where $V_\mu^b$ are the currents of
the global $SU(2)_R$ symmetry, cf.\ Eqs.~(\ref{bcomposite}) and
(\ref{def_uV}). The third component of the current, $V_\mu^3$, the
scalar density and the Abelian field strength are already invariant
under the full gauge-group $SU(2)_L \times U(1)_Y$. The other two
components of the current are only invariant under the non-Abelian
subgroup $SU(2)_L$ but transform non-trivially under the Abelian
group. In order to ensure full $SU(2)_L \times U(1)_Y$ invariance, we
introduced an Abelian phase factor $\phase^{ab}$ coupling to the
charged $SU(2)_R$ currents as given in Eq.~(\ref{def_uV}). The scalar
density and the Abelian field strength excite one-particle states of
the Higgs boson and the photon, respectively, while the currents emit
one-particle states of the massive gauge fields.

Since we couple external sources to gauge-invariant operators only,
the generating functional can be defined in terms of a path integral
without the need to fix a gauge. At tree level, it is given by the
classical action. The equations of motion determine only the physical
degrees of freedom. Hence, they have a whole class of solutions in
terms of the original fields. Every two representatives are related to
each other by a gauge transformation. Since the action is
gauge-invariant, the generating functional is uniquely determined. An
important property of our approach is the fact, that the classical
Goldstone boson field represents the $SU(2)_L$ gauge degrees of
freedom. Thus, no Goldstone boson fields are propagating at the
classical level of the theory. All gauge-invariant sources emit
physical modes only. Moreover, the equations which follow from the
requirement that the variation of the Lagrangian with respect to the
Goldstone boson field vanishes, are not equations of motion but
constraints, expressing the fact that the gauge fields couple to
conserved currents. These constraints can also be obtained by taking
the derivative of the equations of motion for the gauge fields.

The one-loop contribution to the generating functional can be
evaluated with the saddle-point approximation. Because of gauge
invariance, the quadratic form in the path integral representation of
this contribution has zero eigenvalues. They correspond to
fluctuations around the classical fields which are equivalent to
infinitesimal gauge transformations. Hence, the expansion of the
fluctuations involves eigenvectors of the differential operator with
zero and non-zero eigenvalues. In order to evaluate the path integral
one has to equip the space of fields with a metric. The volume element
associated with this metric yields a nontrivial one-loop contribution
to the generating functional. The integration over the zero modes
yields the volume factor of the gauge group, which is absorbed by the
normalization of the integral. The remaining integral over the
non-zero modes is damped by the usual Gaussian factor. It corresponds
to the product of the non-zero eigenvalues of the differential
operator in the quadratic form. As usual, all one-loop contributions
can be expressed in terms of determinants of differential
operators. Since the gauge is not fixed, there are no
ghost contributions in our approach.

As mentioned above, the classical Goldstone boson field represents the
$SU(2)_L$ gauge degrees of freedom. Thus, at the classical level only
physical modes propagate. At the quantum level, however, the situation
is different. Quantum fluctuations around the classical Goldstone
boson field imply virtual Goldstone boson modes propagating within
loops. Note that these modes are absent in any gauge-dependent
approach based on the unitary gauge. They are, however, necessary in
order to ensure a decent high energy behavior of the theory.

The one-loop renormalization of the theory was discussed in detail.
The Green's functions of currents, scalar densities and field
strengths are more singular at short distances than the Green's
functions of the fields. The time ordering of these operators gives
rise to ambiguities which do not occur for the fields
themselves. These ambiguities are reflected by the presence of
additional source terms, which enter Green's functions through contact
terms. We stress that this is a general feature of any field
theory. It is neither particular to our gauge-invariant approach nor
to gauge theories in general. Using dimensional regularization and
employing heat-kernel techniques we analyzed the short distance
properties of the theory. Ultraviolet divergences are related to the
poles of the $d$-dimensional determinant which describes the one-loop
contributions to the generating functional. With this approach we were
able to determine the renormalization prescriptions of the mass
parameter and all coupling constants, fields and source terms,
independently of any renormalization scheme. Due to the dimension of
the source terms involved, the generating functional should be
renormalizable at any loop level. Furthermore, the phase factor which
was introduced in order to ensure full $SU(2)_L \times U(1)_Y$
invariance should not spoil the renormalizability of the theory at any
loop-level either. This is due to the fact that the phase factor only
contains the Abelian gauge degree of freedom which does not affect the
dynamics of the theory.

Finally, we related the bare parameters of the theory to physical
quantities in the on-shell scheme, i.e., we chose the masses of the
Higgs, the $W$- and the $Z$-boson as well as the electric charge as
physical input parameters. The masses were defined as pole positions
of the two-point functions 
\be
\langle 0 | T (\Bphi^\dagger \Bphi)(x) (\Bphi^\dagger \Bphi)(y)| 0
\rangle, \quad 
\langle 0 | T (\phase^{ac} V_\mu^c)(x) (\phase^{bd} V_\nu^d)(y)| 0
\rangle \ .  
\ee
The results for the pole masses can be found in Eqs.~(\ref{MH_pole}),
(\ref{MW_pole}), and (\ref{MZ_pole}). The electric charge was defined
as the residue at the photon pole of the two-point function   
\be  \label{twopoint_B_summary}
\langle 0 | T B_{\mu\nu}(x) B_{\rho\sigma}(y)| 0 \rangle \ . 
\ee
The calculation showed that the result in Eq.~(\ref{e_res_physical})
for the electromagnetic coupling constant defined in this way agrees
with the well known result for the electric charge in the Thompson
limit. This result is quite interesting since there are no Ward
identities in the usual sense between our gauge-invariant Green's
functions. Note, that the usual Ward identities relate gauge-dependent
Green's functions. Hence, in our approach there is ad hoc no identity
relating the residue of the two-point function in
Eq.~(\ref{twopoint_B_summary}) at the photon pole to a three-point
vertex. We did not evaluate any three-point function in order to check
whether it leads to the same result for the coupling constant.

At any rate, the absence of Ward identities does not imply any lack of
knowledge.  All information that Ward identities encode in any
gauge-dependent framework is manifest in our gauge-invariant
approach. As an example this was explicitly discussed for the relation
between the renormalization factors of the coupling constants $g$ and
$\gp$ and those of the gauge fields corresponding to the symmetry
groups $SU(2)_L$ and $U(1)_Y$, respectively.

We have not included fermions in the present analysis of the standard
model. However, the treatment of spin-$1/2$ particles in our approach
is straightforward. One may choose, for instance, the gauge-invariant
fields $\Bphi^\dagger q_L^k, \tBphi^\dagger q_L^k, \Bphi^\dagger
l_L^k$, and $\tBphi^\dagger l_L^k$, cf.\
Eq.~(\ref{Higgs_Fermion}). The corresponding sources which emit
fermi\-on\-ic one-particle states have already been specified in
Ref.~\cite{QED_gaugeinv}. As pointed out in
Refs.~\cite{EW_confinement,tHooft} the complete screening of the
$SU(2)_L$ charge of the composite fields $V_\mu^a, \Bphi^\dagger
\Bphi$ and of the fermionic fields written above can also be
interpreted as the manifestation of confinement in the electroweak
theory, similarly to the mechanism in QCD. As discussed in
Ref.~\cite{tHooft} the physically observed particles then correspond
to ``mesonic'' and ``baryonic'' bound states of the usual fields that
appear in the electroweak Lagrangian, see also
Sec.~\ref{sec:lagrangian}. Our approach, extending the
gauge-invariant treatment to the full group $SU(2)_L \times U(1)_Y$,
can thus be viewed at as a well-defined framework for
carrying out calculations which involve only those external fields
which correspond to the physically observed particles\footnote{Of
course, if we switch on the QCD interactions, the quarks will be
confined in hadrons.}.

We note that the application of our approach to other non-Abelian
gauge theories like QCD is also possible. However, the structure of
the relevant source terms will be different from those used in this
article. The definition of, e.g., $SU(2)_L \times U(1)_Y$ invariant
sources exciting fermionic one-particle states as given in
Ref.~\cite{QED_gaugeinv} is only possible in the spontaneously broken
phase. Hence, an analogous definition does not exist in QCD.
Physically, however, it is not necessary either. Since QCD is
confining, asymptotic states do not carry any $SU(3)_c$ charge. From a
physical point of view it should thus be enough to consider Green's
functions of $SU(3)_c$ invariant operators like, for instance, $\bar
\Psi \Psi$ or $\tr (G_{\mu\nu} G_{\mu\nu})$. Note, however, that the
selection of suitable gauge-invariant Green's functions depends on the
physical problem one wants to investigate. There is no definite choice
which applies to all cases. We believe, however, that a suitable
choice should always be possible, since physical quantities are
gauge-invariant. Any generating functional for QCD that involves
gauge-invariant source terms only can then be evaluated in the
perturbative regime with the same technique as described in this
article. For attempts to dress $SU(3)_c$ charged quarks and gluons
with a non-Abelian generalization of our phase factor see
Refs.~\cite{Lavelle_McMullen,Haller_Chen}.

A first application of our gauge-invariant method can be found in
Ref.~\cite{EWChPT_reanalyzed} where we analyzed the electroweak chiral
Lagrangian~\cite{EW_chiral_Lag}, which describes the low-energy
structure of a strongly interacting electroweak symmetry breaking
sector. In particular, we determined the number of independent
parameters in the effective Lagrangian. Furthermore, we evaluated the
effective Lagrangian for the standard model with a heavy Higgs boson
by matching gauge-invariant Green's functions in the full and the
effective theory. 


\section*{Acknowledgments}

We are grateful to F.~Jegerlehner and V.~Ravindran for enlightening
discussions, a careful reading of the manuscript and suggestions for
improvements.  We are furthermore indebted to J.~Gasser, M.~Knecht,
H.~Leutwyler, E.~de Rafael, J.~Stern, O.~Veretin, and A.~Vicini for
useful discussions. A.N.\ is grateful to the members of the Yale
Physics department for their kind hospitality during the early stages
of this project. He also acknowledges financial support by
Schwei\-zer\-isch\-er Na\-tio\-nal\-fonds during that period.


\appendix 

\section{Differential operators}
\label{app:diffop}

The explicit results for the differential operators which appear in
Sec.~\ref{sec:one_loop} are given below. In the following,
upper case Latin indices $A,B,\ldots$ run from $1$ to $4$, lower case
Latin indices $a,b,\dots$ run from $1$ to $3$, and Greek indices
$\alpha, \beta, \dots$ label the components $1,2$.

The components of the differential operator $\uDfull$ in
Eq.~(\ref{defD}) are given by  
\begin{eqnarray}
d       & = & -\Box + 2 m^2 + 3 m^2 (R^2 - 1) +
                        {1\over 4} \u\Y_\mu^a \u\Y_\mu^a  - \widehat h 
\, , 
                        \label{firstcomp_defD} \\
\delta^b  & = & - \u\Y_\rho^a \u\hvarD_\rho^{ab}
        - {1\over 2} (\u\hvarD_\rho \u\Y_\rho)^b \, , \\
{\delta^T}^a & = &  \u\Y_\rho^a \p_\rho +
        {1\over 2} (\u\hvarD_\rho \u\Y_\rho)^a \, , \\
D^{ab}       & = &   - (\u\hvarD_\rho \u\hvarD_\rho)^{ab} + \delta^{ab}
        \left( m^2 (R^2 - 1)  - \widehat h \right) + M_W^2 R^2
        \delta^{ab} + {1\over 4} \u\Y_\rho^a \u\Y_\rho^b \, , \\ 
\delta_\nu^B & = & M_W R  \u\tY_\mu^A \PTT_{\mu\nu}^{AB}
         \, , \\
\delta_\mu^{T, A} & = & M_W \PTT_{\mu\nu}^{AB} R
\u\tY_\nu^B \, , \\  
\Delta_\nu^{aB} &= & f^{aBc} M_W R \u\Y_\nu^c
        + 2 M_W (\p_\nu R) \delta^{aB} -s M_Z\delta^{4B} 
        \left( 2 \delta^{a3} (\p_\mu R)
        + R T^{ac}\u\W_\mu^c \right) \PT_{\mu\nu} \, ,  \\ 
\Delta_\mu^{T, Ab} & = &  - f^{Abc} M_W R \u\Y_\mu^c
        + 2 M_W (\p_\mu R) \delta^{Ab} + s M_Z \delta^{A4} 
        \PT_{\mu\nu}\left( R \u\W_\nu^c T^{cb}
        - 2 (\p_\nu R) \delta^{3b} \right) \, ,   \\ 
D_{\mu\nu}^{AB} & = & 
        -\delta_{\mu\nu}(\u\tvarD_\rho \u\tvarD_\rho)^{AB} + 2 f^{ABc}
        \uuW^c_{\mu\nu} 
+ (\widetilde M^2)^{AB}\PT_{\mu\nu} + M_W^2 \delta^{AB}\PL_{\mu\nu}
\nonumber \\
        & &\mbox{}+ \PTT_{\mu\alpha}^{AC} (\widetilde
        M^2)^{CD}(R^2-1)\PTT_{\alpha\nu}^{DB} 
        + \delta^{A4} \delta^{B4} \PT_{\mu\rho}
        \widehat J_{\rho\sigma} \PT_{\sigma\nu} \, , 
\label{lastcomp_defD}
\eea 
\noindent
where we introduced the quantities
\bea
\u\varD_\mu^{ab} & = &  \p_\mu\delta^{ab} -
        \varepsilon^{abc}\left( \u\W_\mu^{c}- \delta^{3c}
        \B_\mu^L\right) \, , \\
\u\hvarD_\mu^{ab} & = & \u\varD_\mu^{ab} + {1\over 2} \varepsilon^{abc}
\u\Y_\mu^c  \, , \\
\u\tvarD_\mu^{AB} & = &  \p_\mu\delta^{AB} - f^{ABc} \left(
\u\W_\mu^{c}- \delta^{3c} \B_\mu^L\right) \, , \\
\u\tY_\mu^A               & = & \left( \begin{array}{c}
                                \u\Y_\mu^a \\
                                - {s\over c} \Y_\mu^3 \end{array}
                              \right) \, , \\ 
\uuW_{\mu\nu}^\alpha & = & \p_\mu \u\W_\nu^\alpha - \p_\nu
\u\W_\mu^\alpha - (\W_\mu^3 - B_\mu^L) T_e^{\alpha\beta}
\u\W_\nu^\beta + (\W_\nu^3 - B_\nu^L) T_e^{\alpha\beta} \u\W_\mu^\beta 
\, , \\ 
\uuW_{\mu\nu}^3 & = & \p_\mu \W_\nu^3 - \p_\nu \W_\mu^3 +
\u\W_\mu^\alpha T_e^{\alpha\beta} \u\W_\nu^\beta \, , \\
T_e^{\alpha\beta} & = & \left( \begin{array}{cc}
0  & 1 \\
-1 & 0 
\end{array} \right) \, , \\ 
\PTT_{\mu\nu} &=& {\rm diag}\left(
\delta_{\mu\nu},\delta_{\mu\nu},\delta_{\mu\nu},\PT_{\mu\nu}\right) \, 
, \\
\widetilde M^2 & = & \left( \begin{array}{rrrr}
 M_W^2   &      0 &0 &0 \\
       0  & M_W^2 &0 &0 \\
0 &0 & c^2 M_Z^2 & -c s M_Z^2 \\
0 &0 & -c s M_Z^2 & s^2 M_Z^2 \\
\end{array} \right) \, , \\
\widehat J_{\mu\nu}                 & = & \gps v_{dj}
        \left( \delta_{\mu\nu} J_{\kappa}^\alpha J_{\kappa}^\alpha
        - J_{\mu}^\alpha J_{\nu}^\alpha \right) \ . \label{hatJmunu} 
\eea
Note that the combination $\W_\mu^3 - B_\mu^L$ which appears in the
expressions above is $SU(2)_L \times U(1)_Y$ gauge-invariant, since 
\be
\W_\mu^3 - B_\mu^L = \Z_\mu + s^2 \Z_\mu^T - \A_\mu^T \ . 
\ee

Using the definition from Eq.~(\ref{zeromodes}) of the operator $P$
which creates zero modes, we obtain the following expressions for the
differential operators $\underline{P P^T}$ and $\underline{P^T P}$
which appear in Eq.~(\ref{1loopgf_1}): 
\begin{eqnarray}
\underline{P P^T} & = & \left(\begin{array}{ccc}
0 & 0 & 0 \\
0 & M_W^2 R^2 \delta^{ab}& - M_W R \u\tvarD_\nu^{aB} \\
0 &  M_W \u\tvarD_\mu^{Ab} R & - (\u\tvarD_\mu\u\tvarD_\nu)^{AB}
\end{array}\right) \, , \label{def_PPT} \\ 
\underline{P^T P} & = & \left( \begin{array}{cc}
-\u\varD^{ac}_\mu\u\varD_{\mu}^{cb} + M_W^2 R^2\delta^{ab} & 0 \\
0 & - \Box 
\end{array}\right) .   \label{def_PTP} 
\end{eqnarray}
Furthermore, the operator $\delta_P$ is defined by   
\begin{equation} \label{def_deltaP} 
  \delta_P = {\rm diag}\left( 0, 0,
  \delta^{A4}\delta^{4B}M_W^2\PL_{\mu\nu} \right) \ .
\end{equation} 

The operators $\delta_\mu$ and $\Delta_\mu$ which appear in
Eqs.~(\ref{tildeDmunuW})--(\ref{tildeetamunuA}) are given by the
following expressions: 
\bea
{\delta_\nu^\W}^\alpha & = &  M_W R \u\Y_\nu^\alpha \, ,
\label{delta_nu_W} \\ 
\delta_\nu^\Z & = &  M_Z R \Y_\mu^3 \left( c^2 \delta_{\mu\nu} + s^2
\PT_{\mu\nu}\right) \, , \\  
\delta_\nu^\A & = & s cM_Z R \Y_\mu^3 \PL_{\mu\nu} \, , \\
{\delta_\mu^{\W, T}}^\beta & = & M_W R \u\Y_\mu^\beta \, , \\
\delta_\mu^{\Z, T} & = &  M_Z \left( c^2\delta_{\mu\nu} + s^2
\PT_{\mu\nu}\right) \Y_\nu^3 R \, , \\
\delta_\mu^{\A, T} & = & s c M_Z \PL_{\mu\nu} \Y_\nu^3 R \, , \\
{\Delta_\nu^\W}^{a\beta} &= &  \varepsilon^{a\beta c} M_W R \u\Y^c_\nu 
         + 2 M_W  (\p_\nu R) \delta^{a\beta} \, , \\
{\Delta_\mu^{\W, T}}^{\alpha b} &= &  - \varepsilon^{\alpha bc} M_W R
\u\Y_\mu^c +  2 M_W (\p_\mu R) \delta^{\alpha b} \, , \\
{\Delta_\nu^\Z}^{a} &= &   \varepsilon^{a3c} M_Z R
        (c^2 \u\Y_\nu^c - s^2 \u\W_\mu^c \PT_{\mu\nu}) + 2
\delta^{a3}M_Z (\p_\mu R)\left(c^2\delta_{\mu\nu} + s^2
        \PT_{\mu\nu}\right)  \, , \\
{\Delta_\nu^\A}^{a} &= &   \varepsilon^{a3c} s c M_Z R
        (\u\Y_\nu^c + \u\W_\mu^c \PT_{\mu\nu}) + 2 s c \delta^{a3} M_Z
(\p_\mu R) \PL_{\mu\nu} \, ,  \\
{\Delta_\mu^{\Z, T}}^{b} &= &  - \varepsilon^{3bc} M_Z
        (c^2 \u\Y_\mu^c - s^2 \PT_{\mu\nu}\u\W_\nu^c ) R + 2
\delta^{3b} M_Z \left(c^2\delta_{\mu\nu} + s^2 
        \PT_{\mu\nu}\right) (\p_\nu R) \, ,   \\
{\Delta_\mu^{\A, T}}^{b} &= &   - \varepsilon^{3bc} s c M_Z
        (\u\Y_\mu^c + \PT_{\mu\nu} \u\W_\nu^c ) R + 2 s c \delta^{3b}
M_Z \PL_{\mu\nu} (\p_\nu R) \ .
\label{Delta_mu_A_T}
\eea 
The first terms on the right-hand side of
Eqs.~(\ref{tildeDmunuW})--(\ref{tildeetamunuA}) read as follows:   
\bea
{D^\W_{\mu\nu}}^{\alpha\beta} & = & {D^{W,0}_{\mu\nu}}^{\alpha\beta}
        + T_e^{\alpha\beta} \left((\p_\sigma(\W_\sigma^3-B_\sigma^L)) 
        + 2(\W_\sigma^3-B_\sigma^L)\p_\sigma\right)\delta_{\mu\nu}
-(T_e^{\alpha\gamma}\u\W_\sigma^\gamma)
        (\u\W_\sigma^\delta T_e^{\delta\beta})\delta_{\mu\nu} 
\nonumber \\
&&+ (\W_\sigma^3-B_\sigma^L) (\W_\sigma^3-B_\sigma^L)
        \delta_{\mu\nu} \delta^{\alpha\beta}  
+ 2 T_e^{\alpha\beta} \uuW_{\mu\nu}^3 + M_W^2 (R^2-1)
    \delta_{\mu\nu} \delta^{\alpha\beta} \, ,
\label{firstcomp_tildevarD} \\ 
D^\Z_{\mu\nu} & = & D^{\Z,0}_{\mu\nu}  + c^2
        \u\W_\sigma^\alpha\u\W_\sigma^\alpha\delta_{\mu\nu}
        + s^2 \PT_{\mu\rho}\widehat J_{\rho\sigma} \PT_{\sigma\nu}
\nonumber \\ 
&&+ M_Z^2   \left( c^2 \delta_{\mu\rho} + s^2 \PT_{\mu\rho}\right)
  (R^2-1)   \left( c^2 \delta_{\rho\nu} + s^2 \PT_{\rho\nu}\right)
\, , \\
D^\A_{\mu\nu} & = & D^{\A,0}_{\mu\nu}  + s^2
\u\W_\sigma^\alpha\u\W_\sigma^\alpha\delta_{\mu\nu}  
+ s^2 M_W^2 \PL_{\mu\rho} (R^2-1) \PL_{\rho\nu} 
+ c^2 \PT_{\mu\rho}\widehat J_{\rho\sigma} \PT_{\sigma\nu} \, , \\
{\xi^\Z_{\mu\nu}}^\alpha & = &  - c T_e^{\alpha\beta}
\left((\p_\sigma\u\W_\sigma^\beta) + 
        2 \u\W_\sigma^\beta \p_\sigma\right) \delta_{\mu\nu} 
        - c (\W_\sigma^3-B_\sigma^L) \u\W_\sigma^\alpha \delta_{\mu\nu}
        - 2 c T_e^{\alpha\beta} \uuW_{\mu\nu}^\beta \, , \\
{\xi^{\Z, T}_{\mu\nu}}^\beta & = &  - c
\left((\p_\sigma\u\W_\sigma^\alpha) + 
        2 \u\W_\sigma^\alpha \p_\sigma\right) T_e^{\alpha\beta}
        \delta_{\mu\nu}  
        - c (\W_\sigma^3-B_\sigma^L) \u\W_\sigma^\beta \delta_{\mu\nu}
        - 2c  \uuW_{\mu\nu}^\alpha T_e^{\alpha\beta} \, , \\
{\xi^\A_{\mu\nu}}^\alpha & = & - s T_e^{\alpha\beta}
\left((\p_\sigma\u\W_\sigma^\beta) + 
        2\u\W_\sigma^\beta\p_\sigma\right) \delta_{\mu\nu} 
        - s (\W_\sigma^3-B_\sigma^L) \u\W_\sigma^\alpha
        \delta_{\mu\nu} 
        - 2 s T_e^{\alpha\beta} \uuW_{\mu\nu}^\beta \, , \\
{\xi^{\A, T}_{\mu\nu}}^\beta & = & - s
\left((\p_\sigma\u\W_\sigma^\alpha) + 
                         2\u\W_\sigma^\alpha\p_\sigma\right)
                         T_e^{\alpha\beta} \delta_{\mu\nu}
        - s (\W_\sigma^3-B_\sigma^L) \u\W_\sigma^\beta \delta_{\mu\nu}
        - 2s  \uuW_{\mu\nu}^\alpha T_e^{\alpha\beta} \, , \\
\eta^\A_{\mu\nu} & = & sc 
\u\W_\sigma^\alpha \u\W_\sigma^\alpha\delta_{\mu\nu}  
+ s c M_Z^2   \left( c^2 \delta_{\mu\rho} + s^2 \PT_{\mu\rho}\right)
        (R^2-1) \PL_{\rho\nu} 
         - s c \PT_{\mu\rho}\widehat J_{\rho\sigma} \PT_{\sigma\nu} \, 
, \\
\eta^{\A, T}_{\mu\nu} & = & sc
        \u\W_\sigma^\alpha \u\W_\sigma^\alpha\delta_{\mu\nu} 
+ s c M_Z^2 \PL_{\mu\rho}(R^2-1) 
  \left( c^2 \delta_{\rho\nu} + s^2 \PT_{\rho\nu}\right) - s c
\PT_{\mu\rho}\widehat J_{\rho\sigma} 
\PT_{\sigma\nu} \ . \label{lastcomp_tildevarD} 
\end{eqnarray}

Free propagators are the inverse of the following operators: 
\begin{eqnarray}
d^0 & \equiv & d_m = -\Box + 2 m^2 \, , \label{D_H_free} \\
{\Theta^0}^{ab} & \equiv & {D^0}^{ab} =  (-\Box + M_W^2) \delta^{ab}
\, , \\
{\varD^{\W,0}_{\mu\nu}}^{\alpha\beta} & \equiv &
{D^{\W,0}_{\mu\nu}}^{\alpha\beta} = 
(-\Box + M_W^2)\delta_{\mu\nu} \delta^{\alpha\beta} \, , \\
\varD^{\Z,0}_{\mu\nu} & \equiv & D^{\Z,0}_{\mu\nu} = 
(-\Box  + M_Z^2) \PT_{\mu\nu} + (-\Box + M_W^2)\PL_{\mu\nu} \, , \\
\varD^{\A,0}_{\mu\nu} & \equiv & D^{\A,0}_{\mu\nu}  =  
-\Box\PT_{\mu\nu} + (-\Box + M_W^2)\PL_{\mu\nu} \, ,  \label{D_A_free}
\end{eqnarray} 
which are obtained in the limit where all sources are switched off. We
observe that all transversal components propagate with the proper
mass, while all longitudinal components propagate with the $W$-boson
mass.

Since we perform a saddle-point approximation in the path integral,
the fields which appear in the list of differential operators in
Eqs.~(\ref{firstcomp_defD})--(\ref{lastcomp_tildevarD}) obey the
equations of motion (\ref{eomRcomp})--(\ref{eomUcompZ}). We have
used this fact to simplify the expressions of those operators in
Eqs.~(\ref{firstcomp_defD})--(\ref{lastcomp_defD}) which correspond
to the fluctuations $\eta^a$ of the Goldstone bosons.  Furthermore, it
is important to ensure that the full differential operator $\uDfull$
is Hermitian, i.e.\ satisfies the relation $(y,[\uDfull] y^\prime) =
(y^\prime,[\uDfull] y)$ for arbitrary fluctuation vectors
$y,y^\prime$.


\section{Renormalization prescriptions}
\label{app:renorm_prescriptions}

For the calculation of the counterterms and the renormalization
prescriptions it is convenient to switch to a real 
$O(4)$-representation for the Higgs field: 
\be
  \phi = \left( \begin{array}{c}
                        \phi^1 \\
                        \phi^2 \\
                        \phi^3 \\
                        \phi^4
                \end{array} \right) \ . 
\ee
The covariant derivative for the Higgs field is given by 
\bea
\nabla_\mu^{NM} \phi^M & = & (\partial_\mu\delta^{NM} + F_\mu^{NM})
\phi^M \ , \quad N,M = 1,2,3,4 \, ,  \\ 
F_\mu & = & W_\mu^a T_L^a + B_\mu T_R^3, \quad a = 1,2,3 \ , 
\eea
where the matrices $T_{L,R}$ are defined through  
\bea
T_R^a & = & \half (\vT^a + \aT^a ) , \quad   
T_L^a = \half (\vT^a - \aT^a ) \, , \nonumber \\
(\vT_c)^{NM} & \doteq & - \varepsilon^{NMc} , \quad  
(\aT_c)^{NM} \doteq \delta^N_4 \delta^M_c - \delta^M_4 \delta^N_c
\, . 
\eea
It can easily be shown that the matrices $T_R^a$ and $T_L^a$
separately satisfy the Pauli algebra. The Lagrangian in
Eq.~(\ref{lagSMfull}) can then be rewritten in terms of the fields in
the real representation. The main changes compared to the notation
with a complex doublet $\Bphi$ are obtained by the following
replacements:
\bea
{1\over2} D_\mu\Bphi^\dagger D_\mu\Bphi & \to & 
{1\over2} \nabla_\mu\phi^T\nabla_\mu\phi \quad , \quad 
\Bphi^\dagger \Bphi \to \phi^T\phi \, , \nonumber \\
J_\mu^a \phase^{ab} V_\mu^b & \to & 
4 J_\mu^a \phase^{ab} \JRmu^b \, , 
\eea
with 
\be
\JRmu^a =  \phi^T T_R^a \nabla_\mu\phi . 
\ee

The evaluation of the path integral representation of the generating
functional in Eq.~(\ref{genfunc_saddlepoint}) is again performed by a
saddle-point approximation around the classical action. The quantum
fluctuations are introduced as simple shifts in the fields
\bea
\phi^N & \to & \phi^N + f^N  \, , \nonumber \\   
F_\mu & \to & F_\mu + g \w_\mu^a T_R^a + g' \b_\mu T_L^3 \, .      
\eea
It will be useful to treat the fluctuations $\w_\mu^a, \b_\mu$ and
the matrices $T_L^a, T_R^3$ in a unified way by introducing the
quantities: 
\be
\tq_\mu^A \doteq \left( \begin{array}{c}
\w_\mu^a \\ \b_\mu \end{array} \right) , \qquad
t^A \doteq \left( \begin{array}{c} 
T_L^a \\ T_R^3 \end{array} \right) , \qquad 
A=1,2,3,4 . 
\ee

The terms quadratic in the fluctuations determine the differential 
operator $\Dfull$, after the proper treatment of the zero modes. From
the path integral measure one gets the operator $\bP^T \bP$. The final
result for these operators can be written in the following way, which
is suitable for using the heat-kernel method and the
short distance expansion outlined in Sec.~\ref{sec:renorm},   
\be
\Dfull = \Dloc + \dDnl \, , 
\ee
where the local part $\Dloc$ of the differential operator is given by
\bea
\Dloc & = & - \bD_\rho \bD_\rho + \bsigma \, , \label{Dloc} \\
\bD_\rho & = & \left( \begin{array}{cc}
\bd_\rho & 0 \\
0 & \delta_{\mu\nu} \bvarD_\rho   \end{array} \right) \, , \\
\bd_\rho^{NM} & = & \delta^{NM} \p_\rho + F_\rho^{NM} - 4
j_\rho^{NM} \, , \\ 
\bvarD_\rho^{AB} & = & \delta^{AB} \p_\rho - f^{ABc}
W_\rho^c \, , \\
\bsigma & = & \left( \begin{array}{cc}
\bsigma_{ff} & \bsigma_{fq} \\ 
\bsigma_{qf} & \bsigma_{qq} \end{array} \right) \, , \\
\bsigma_{ff}^{NM} & = & \left( - m^2 + \lambda \phi^T \phi - h -
v_{jj} J_\mu^\alpha J_\mu^\alpha - c_{jj} J_\mu^Z J_\mu^Z \right)
\delta^{NM} \nonumber \\ 
& &\mbox{}+ 2 \lambda \phi^N \phi^M 
- (t^A \phi \otimes \phi^T t^A)^{NM} + 16 (j_\rho j_\rho)^{NM} \, , \\ 
\bsigma_{fq,\nu}^{NB} & = & - 2 (t^B \nabla_\nu \phi)^N + 4 \{ j_\nu
, t^B \}^{NM} \phi^M \, , \\
\bsigma_{qf,\mu}^{AM} & = & 2 (\nabla_\mu \phi^T t^A)^M + 4 \phi^N \{
j_\mu , t^A \}^{NM} \, , \\
\bsigma_{qq,\mu\nu}^{AB} & = & - \delta_{\mu\nu} \phi^T t^A t^B \phi -
2 {W_{\mu\nu}^{Adj}}^{AB} + \delta^{A4} \delta^{B4} \widehat
J_{\mu\nu} \, . 
\eea
Here we introduced the quantities 
\bea
j_\mu & = & j_\mu^a T_R^a \, , \\  
{W_{\mu\nu}^{Adj}}^{AB} & = & \left[ \bvarD_\mu , \bvarD_\nu \right]^{AB}
= - f^{ABc} W_{\mu\nu}^c \ .   
\eea
The quantities $f^{ABc}$ and $\widehat J_{\mu\nu}$ have been defined in
Eqs.~(\ref{def_tvarD}) and (\ref{hatJmunu}), respectively. 

The nonlocal part $\dDnl$ of the differential operator which contains
the contributions from the phase factor reads 
\bea
\dDnl & = & \left( \begin{array}{cc}
0 & \dDnl_{fq} \\
\dDnl_{qf} & \dDnl_{qq}  \end{array} \right) \, , \label{dDnl} \\
\dDnl_{fq,\nu}^{NB} & = & - 4 \gp \delta^{B4} \left( 2 j_\rho^\alpha
T_e^{\alpha\beta} T_R^\beta \nabla_\rho \phi + (d_\rho j_\rho)^\alpha
T_e^{\alpha\beta} 
T_R^\beta \phi \right)^N \left( \p_\nu {1\over \Box} \right) 
- 4 \gp \delta^{B4} \left( j_\rho^\alpha T_e^{\alpha\beta}
T_R^\beta \phi \right)^N \left( \p_\rho \p_\nu {1\over \Box} \right)
\, , \\
\dDnl_{qf,\mu}^{AM} & = & 4 \gp \delta^{A4} \left( \p_\mu {1\over
  \Box} \right) \left( \phi^T j_\rho^\alpha T_e^{\alpha\beta}
T_R^\beta \nabla_\rho - (\nabla_\mu \phi)^T j_\rho^\alpha T_e^{\alpha\beta}
T_R^\beta \right)^M \, , \\
\dDnl_{qq,\mu\nu}^{AB} & = & 8 \gp \delta^{A4} \left( \p_\mu {1\over
    \Box} \right) \left( \phi^T j_\nu^\alpha T_e^{\alpha\beta}
  T_R^\beta t^B \phi \right) 
+ 4 \gps \delta^{A4} \delta^{B4} \left( \p_\mu {1\over \Box}
\right) \left( \phi^T j_\rho^\alpha T_R^\alpha \nabla_\rho \phi
\right) \left( \p_\nu {1\over \Box} \right) \nonumber \\ 
& &\mbox{}- \delta^{A4} \delta^{B4} \left( \p_\mu \p_\sigma {1\over
  \Box} \right) \widehat J_{\sigma\nu} 
- \delta^{A4} \delta^{B4} \widehat J_{\mu\sigma} \left(
\p_\sigma \p_\nu {1\over \Box} \right) 
+ \delta^{A4} \delta^{B4} \left( \p_\mu \p_\sigma {1\over \Box}
\right) \widehat J_{\sigma\kappa} \left( \p_\kappa \p_\nu {1\over \Box}
\right)  \ . 
\eea

The contributions to the counterterms from the path integral measure
can be calculated from the expression  
\be
(\bP^T \bP)^{AB}  =  - (\bvarD_\rho \bvarD_\rho)^{AB} -
\phi^T t^A t^B \phi \ . \label{bPTbP}
\ee

The determination of the counterterms then proceeds along the lines
sketched in Sec.~\ref{sec:renorm}. From the counterterms we can
read off the renormalization prescriptions for the fields, the mass
parameter $m^2$, the coupling constants and the sources. The relations
between bare and renormalized fields, masses and coupling constants
which are needed in Sec.~\ref{sec:greensfunctions} are given by
\bea
W_\mu^a & = & W_\mu^{a,r} \, , \label{W_mu_ren} \\
B_\mu   & = & B_\mu^r \, , \label{B_mu_ren} \\
\phi    & = & Z_\phi^{1/2} \phi_r \, , \\
Z_\phi  & = & 1 - (6 \grs + 2 \gprs) (\polemr + \delta z) \, , 
\label{Zphi_ren} \\
m^2     & = & \mrs \left[ 1 - {1\over2} (24\lr + 3\grs + \gprs) 
        (\polemr + \delta m^2) - (Z_\phi - 1) \right] \, , \label{m_ren}
\\
\lambda & = & \lr \Bigg[ 1 - \left( 24 \lr + 3 \grs + \gprs + 
        {3\over8} {(\grs + \gprs)^2 + 2 \gr^4 \over\lr} \right)
        (\polemr + \delta \lambda) 
        - 2 (Z_\phi - 1) \Bigg] \, , \\
g^2     & = & \grs \left[ 1 + {43\over3} \grs 
        (\polemr + \delta g^2) \right] \, , \\
\gp^2   & = & \gprs \left[ 1 - {1\over3} \gprs (\polemr + \delta \gps) 
        \right] \, , \label{gprime_ren} 
\eea
where we denoted the pole term by 
\be
\polemr   \doteq {\mu^{d-4} \over 16\pi^2} \left( {1 \over d-4} -
{1\over2} (\ln(4\pi) +    \Gamma'(1) + 1) \right) 
+ {1\over 32\pi^2} \ln \left( {2\mrs \over \mu^2} \right) \ . 
\ee

The finite renormalization constants $\delta m^2, \ldots , 
\delta \gps$ which appear in the
Eqs.~(\ref{m_ren})--(\ref{gprime_ren}) are determined by the
renormalization conditions given in Sec.~\ref{sec:greensfunctions}.

The renormalization of the source terms can be achieved by using the
following prescriptions: 
\begin{eqnarray}
h       & = & c_h h_r \, , \label{firstsource_ren} \\
c_h     & = & 1 - {1\over2} (24\lr + 3\grs + \gprs)
        (\polemr + \delta c_h) - (Z_\phi - 1) \, , 
        \\ 
K_{\mu\nu} & = & K_{\mu\nu}^r \, , \\ 
J_{\mu}^\alpha & = & c_v J_{\mu}^{\alpha,r} \, , \\ 
c_v     & = & 1 - {1\over4} (24\grs + 2\gprs)
        (\polemr + \delta c_v) - (Z_\phi - 1) \, , \\
J_{\mu}^Z & = & c_Z J_{\mu}^{Zr} \, , \\
c_Z     & = & 1 - (6\grs + 2\gprs)
        (\polemr + \delta c_Z) - (Z_\phi - 1) \, , \\ 
v_{jj}  & = & v_{jj}^r + \left( 24\grs + 2\gprs 
        + {9\over8} v_{dj}^r \gpr^4 - {1\over2} (v_{jj}^r +4) 
        (24\lr + 3\grs + \gprs) \right) (\polemr + \delta v_{jj})
\nonumber \\ 
&& - v_{jj}^r (Z_\phi - 1) - 2 v_{jj}^r (c_v - 1) \, , \\
c_{jj}  & = & c_{jj}^r + \left( 24\grs + 8 \gprs 
        -{1\over2} (c_{jj}^r+4) 
        (24\lr + 3\grs + \gprs) \right) (\polemr + \delta c_{jj})
- c_{jj}^r (Z_\phi - 1) - 2 c_{jj}^r (c_Z - 1) \, , \\
c_{Bj}  & = & c_{Bj}^r + {4\over 3} (\polemr + \delta c_{Bj}) 
        - c_{Bj}^r (c_Z - 1) \, , \\
c_{Bjj} & = & c_{Bjj}^r - {16\over 3} (\polemr + \delta c_{Bjj}) 
        - 2 c_{Bjj}^r (c_v - 1) \, , \\
v_{djj} & = & v_{djj}^r - {128\over3} (\polemr + \delta v_{djj}) 
        - v_{djj}^r (c_Z - 1) - 2 v_{djj}^r (c_v - 1) \, , \\
v_{dj}  & = & v_{dj}^r + {16\over3} (\polemr + \delta v_{dj}) 
        - 2 v_{dj}^r (c_v - 1) \, , \\
c_{djj} & = & c_{djj}^r - {128\over3} (\polemr + \delta c_{djj}) 
        - c_{djj}^r (c_Z - 1) - 2 c_{djj}^r (c_v - 1) \, , \\
c_{dj}  & = & c_{dj}^r + {16\over3} (\polemr + \delta c_{dj}) 
        - 2 c_{dj}^r (c_Z - 1) \, , \\
v_{JJ2} & = & v_{JJ2}^r - \left( {37\over48} (v_{dj}^r)^2 \gpr^4 
        + 2 (v_{jj}^r +4)^2 - {64\over3}\right) (\polemr + \delta v_{JJ2}) 
        - 4 v_{JJ2}^r (c_v - 1) \, , \\
v_{JJJJ} & = & v_{JJJJ}^r - \left( {7\over24} (v_{dj}^r)^2 \gpr^4 
        + {64\over3}\right) (\polemr + \delta v_{JJJJ})  
        - 4 v_{JJJJ}^r (c_v - 1) \, , \\
c_{JJ2} & = & c_{JJ2}^r - 2 (c_{jj}^r + 4)^2 (\polemr + \delta c_{JJ2}) 
        - 4 c_{JJ2}^r (c_Z - 1) \, , \\
v_{J2ZZ} & = & v_{J2ZZ}^r - \left( 4 (v_{jj}^r + 4) 
        (c_{jj}^r + 4) - {128\over3} \right) (\polemr + \delta v_{J2ZZ}) 
- 2 v_{J2ZZ}^r (c_v - 1) - 2 v_{J2ZZ}^r (c_Z - 1) \, , \\
v_{JJZZ} & = & v_{JJZZ}^r - {128\over3} (\polemr + \delta v_{JJZZ}) 
        - 2 v_{JJZZ}^r (c_v - 1) 
        - 2 v_{JJZZ}^r (c_Z - 1) \, , \\
c_{hh}  & = & c_{hh}^r - 2 (\polemr + \delta c_{hh}) - 2 c_{hh}^r (c_h
- 1) \, , \\ 
c_{mh}  & = & c_{mh}^r - 4 (\polemr + \delta c_{mh}) 
        - c_{mh}^r (c_h - 1) 
        - c_{mh}^r \left({m^2 - \mrs \over \mrs}\right) \, , \\
c_{hJJ} & = & c_{hJJ}^r - 4 (v_{jj}^r + 4) (\polemr + \delta c_{hJJ}) 
        - c_{hJJ}^r (c_h - 1) 
        - 2 c_{hJJ}^r (c_v - 1) \, , \\   
c_{hZZ} & = & c_{hZZ}^r - 4 (c_{jj}^r + 4) (\polemr + \delta c_{hZZ}) 
        - c_{hZZ}^r (c_h - 1) 
        - 2 c_{hZZ}^r (c_Z - 1) \, , \\   
c_{mJJ} & = & c_{mJJ}^r - 4 (v_{jj}^r + 4) (\polemr + \delta c_{mJJ}) 
        - 2 c_{mJJ}^r (c_v - 1) 
        - c_{mJJ}^r \left({m^2 - \mrs \over \mrs}  \right) \, , \\
c_{mZZ} & = & c_{mZZ}^r - 4 (c_{jj}^r + 4) (\polemr + \delta c_{mZZ}) 
        - 2 c_{mZZ}^r (c_Z - 1)  
        - c_{mZZ}^r \left({m^2 - \mrs \over \mrs}  \right) \, . 
\label{lastrenormprescription}
\end{eqnarray}
In order to renormalize on-shell quantities, like masses, coupling
constants and $S$-matrix elements it will not be necessary to
determine all the finite renormalization constants $\delta c$ and
$\delta v$ which appear in
Eqs.~(\ref{firstsource_ren})--(\ref{lastrenormprescription}).


\section{Self-energies}
\label{app:selfenergies}

The explicit results for the self-energies which appear in the
generating functional in Eq.~(\ref{W_1loop_sources}) are given by the  
following expressions:
\bea
\Sigma_{\Bphi^\dagger\Bphi}(p^2) & = & {e^2 \over s^2 c^2} \Bigg\{ 
\left( {3\over4} {2m^2\over M_Z^2} \right) A_0(2m^2) 
+ \left({1\over 4} {p^2 \over M_Z^2} + {3\over 4} {2m^2\over 
M_Z^2} + {13\over4} {M_W^2 \over M_Z^2} 
- {1\over 4} \right) A_0(M_W^2) 
\nonumber \\
&&\mbox{}
+ \left( - {1\over 4} {p^2 \over M_Z^2} + {7\over 4} 
- {1\over 4} {M_W^2\over M_Z^2} \right) A_0(M_Z^2) 
\nonumber \\
&&\mbox{}+ \left( - {1\over 4} s^2 c^2 M_Z^2 \right) B_0(0,M_W^2;0) 
+ \left( {1\over 4} s^2 M_Z^2 \right) B_0(M_Z^2,0;0) 
\nonumber \\
&&\mbox{}+ \Bigg( - {1\over 8} {p^4\over M_Z^2} + {5\over 2} {p^2
M_W^2\over M_Z^2} + {3\over 8} {(2m^2)^2\over M_Z^2} 
+ {3\over 2} {2m^2 M_W^2\over M_Z^2}  
+ 3 {M_W^4\over M_Z^2} \Bigg) B_0(M_W^2,M_W^2;p^2) 
\nonumber \\
&&\mbox{}+ \left( {1\over 8} {p^4\over M_Z^2} + {1\over 2} p^2  +
{3\over 2} M_Z^2 \right) B_0(M_Z^2,M_Z^2;p^2)
+ \left( {9\over 8} {(2m^2)^2\over M_Z^2} \right)
B_0(2m^2,2m^2;p^2) 
\Bigg\} \, , 
\label{def_Sigma_H} \\
\Sigma_\uW^T(p^2) & = & {e^2 \over s^2 c^2} \Bigg\{ 
\left( - {c^2 \over 12} {2m^2 - M_W^2 \over p^2} + {7\over 12} c^2
\right) A_0(2m^2) \nonumber \\ 
&&\mbox{}+\left( {c^2 2m^2 + M_W^2 + 6 c^2 M_W^2 - 8 c^4
M_W^2 \over 12 p^2} + {2 \over 3} {c^2 p^2 \over M_W^2} 
+ 3 c^2 {M_W^2 \over 2m^2} - {11\over 6} c^2 + {4\over 3} c^4 \right)
A_0(M_W^2) \nonumber \\  
&&\mbox{}+ \Bigg( \left[ -1 - 7 c^2 + 4 c^4 + 4 c^6\right] {M_W^2
\over 12 p^2} - {2\over 3} {c^4 p^2 \over M_Z^2} + {3\over 2} {M_W^2
\over 2m^2}  
+ {1\over 12} c^2 - {5\over 3} c^4 - {1\over 3} c^6
\Bigg) A_0(M_Z^2) \nonumber \\   
&&\mbox{}+ \left( - {2\over 3} {c^2 p^4 \over M_W^2} + {5\over 3} c^2
p^2 + {7\over 3} c^2 M_W^2 \right) B_0(M_W^2, M_W^2; p^2) \nonumber \\ 
&&\mbox{}+ \Bigg( \left[- 1 - 6 c^2 + 11 c^4 - 4 c^8\right] {M_W^4\over
12 c^2 p^2} + {2\over 3} {c^6 p^4 \over M_W^2}  
+ \left[ - {1\over 12} c^2 + {14\over 3} c^4 + c^6 \right] p^2
\nonumber \\ 
&&\qquad\mbox{}  + \left[ {5 \over 6} - {29\over 6}
c^2 + {2\over 3} c^4 \right] M_W^2   \Bigg) 
B_0(M_W^2, M_Z^2; p^2) \nonumber \\  
&&\mbox{}+ \Bigg( - {c^2 \over 12} {(2m^2 - M_W^2)^2 \over p^2} -
{1\over 12} c^2 p^2  
- {1\over 6} c^2 2m^2 + {5\over 6} c^2 M_W^2
\Bigg) B_0(M_W^2, 2m^2; p^2) \nonumber \\  
&&\mbox{}+ \Bigg( - {s^2 c^4 M_W^4 \over 3 p^2} + (1 - c^4) {2 c^2 p^4
\over 3 M_W^2}  + \left[ {19\over 3} c^2 - {16\over 3} c^4 - c^6
\right] p^2 
+ \left[ - {1\over 3} c^2 + {1\over 3} c^4
\right] M_W^2 \Bigg) B_0(0, M_W^2; p^2) \nonumber \\   
&&\mbox{}+ {1\over 16 \pi^2} \Bigg( \left[ - {11\over 9} c^2 +
{4\over 3} c^4 \right] p^2 - {1\over 6} c^2 2m^2  
+ \left[ {1\over
c^2} + 2 c^2 \right] {M_W^4 \over 2m^2} 
+ \left[ - {1\over 6} - {7\over 3} c^2 - {2\over 3} c^4 \right] M_W^2
\Bigg) \Bigg\} \, , 
\label{def_SigmaT_W} \\
\Sigma_\uW^L(p^2) & = & 
\left( \half {M_W^2 - 2m^2 \over p^2} - {3\over 2} \right) A_0(2m^2)
+ \left( \left[3 - 4 c^2 + {1\over 2c^2}\right] {M_W^2 \over
p^2} + \half {2m^2 \over p^2} - 6 {M_W^2 \over 2m^2} - 2 \right) A_0(M_W^2)
\nonumber \\  
&&\mbox{}+ \left( \left[- \half - {7\over 2} c^2 + 2 c^4 + 2
c^6\right] {M_Z^2 \over p^2} + \half - 2 c^2 + 2 c^4 \right)
A_0(M_Z^2) \nonumber \\   
&&\mbox{}+ \left( \half p^2 - 2 M_W^2 \right) B_0(M_W^2, M_W^2; p^2)
\nonumber \\  
&&\mbox{}+ \Bigg( \left[- {1\over 2c^4} - {3 \over c^2} + {11\over 2} - 2
c^4 \right] {M_W^4 \over p^2}  
+ \left[ - \half + 2 c^2 - 2 c^4\right] p^2 
+ \left[ -{1\over  
c^2} + 3 - 4 c^4\right] M_W^2 \Bigg) B_0(M_W^2, M_Z^2; p^2) \nonumber \\ 
&&\mbox{}+ \left( - \half {(2m^2 - M_W^2)^2 \over p^2} - 2 M_W^2 \right)
B_0(M_W^2, 2m^2; p^2) \nonumber \\  
&&\mbox{}+ \Bigg( - {2 s^2 c^2 M_W^4 \over p^2} - 2 s^2 c^2 p^2
+ \left[ - 6 + 2 c^2 + 4 c^4\right] M_W^2 \Bigg) B_0(0, M_W^2; p^2)
\nonumber \\  
&&\mbox{}
+ {1\over 16 \pi^2} \left( \left[ -4 - {2\over c^4}\right]
{M_W^4 \over 2m^2} + 4 s^2 M_W^2 \right) \, , 
\label{def_SigmaL_W} \\
\Sigma_\Z^T(p^2) & = & {e^2 \over s^2 c^2} \Bigg\{
\left( - {1\over 12} {2m^2 - M_Z^2 \over p^2} + {7\over 12} \right)
A_0(2m^2)  
+ \left( 3 {M_W^2 \over 2m^2} + {1\over 6} + {4\over 3} c^2 -
4 c^4 \right) A_0(M_W^2) \nonumber \\
&&\mbox{}
+ \left( {1\over 12} {2m^2 - M_Z^2 \over p^2} + {3\over 2} {M_Z^2
\over 2m^2} + {1\over 12} \right) A_0(M_Z^2) \nonumber \\  
&&\mbox{}+ \Bigg( \left[ - {1\over 12} + {1\over 3} c^2 + 7 c^4 \right] p^2
+ \left[{5\over 3} + {4\over 3} c^2 - 4 c^4 \right]
M_W^2 \Bigg) B_0(M_W^2, M_W^2; p^2) \nonumber \\
&&\mbox{}+ \Bigg( - {1\over 12} {(2m^2 - M_Z^2)^2 \over p^2} 
- {1\over 12} p^2 - {1\over 6} 2m^2 + {5\over 6} M_Z^2 \Bigg)
B_0(M_Z^2, 2m^2; p^2) \nonumber \\ 
&&\mbox{}+ {1\over 16 \pi^2} \Bigg( \left[ - {1\over 9} + {2\over 9} c^2
\right] p^2 - {1\over 6} 2m^2 + \left[ 1 + 2 c^4 \right] {M_Z^4
\over 2m^2}  
+  \left[- {1\over 6} - {1\over 3} c^2 +
{4\over 3} c^4 - 4 c^6 \right] M_Z^2 \Bigg)
\Bigg\} \, , 
\label{def_SigmaT_Z} \\ 
\Sigma_\Z^L(p^2) & = & 
\left( - \half {2m^2 - M_Z^2\over p^2} - {3\over 2}
\right) A_0(2m^2)  
+ \left( - 6 {M_W^2 \over 2m^2} - 2 \right) A_0(M_W^2)  
+ \left( \half {2m^2 - M_Z^2 \over p^2} - 3 {M_Z^2
\over 2m^2} + \half \right) A_0(M_Z^2) \nonumber \\
&&\mbox{}+ \left( - 4 M_W^2 \right) B_0(M_W^2, M_W^2; p^2) 
+ \left( \half p^2 + 2m^2 - M_W^2 \right) B_0(M_W^2, 2m^2; p^2) \nonumber \\ 
&&\mbox{}+ \left( -\half {(2m^2 - M_Z^2)^2 \over p^2} - \half p^2 - 2m^2 -
M_Z^2 \right) B_0(M_Z^2, 2m^2; p^2)  
+ {1\over 16 \pi^2} \left( - 4 {M_W^4 \over 2m^2} - 2 {M_Z^4
\over 2m^2} \right)  \, , 
\label{def_SigmaL_Z} \\ 
\Sigma_{\Z\A}^T(p^2) & = & {e^2 \over s c} \Bigg\{
\left( - {2\over 3} + 4 c^2 \right) A_0(M_W^2) 
+ \left( \left[ - {2\over 3} + 4 c^2 \right] M_W^2 + \left[ -
{1\over 6} - 7 c^2 \right] p^2 \right) B_0(M_W^2,M_W^2;p^2) \nonumber \\
&&\mbox{}+ {1\over 16 \pi^2} \left( - {2\over 3} M_W^2 + 4 c^2 M_W^2 -
{1\over 9} p^2 \right)
\Bigg\} \, , 
\label{def_SigmaT_ZA} \\
\Sigma_{\A\Z}^T(p^2) & = & \Sigma_{\Z\A}^T(p^2) \, , \\
\Sigma_{\Z\A}^L(p^2) & \equiv & 0 \, , \label{def_SigmaL_ZA} \\ 
\Sigma_\A^T(p^2) & = & - e^2 \Bigg\{
4 A_0(M_W^2) + \left( 4 M_W^2 - 7 p^2 \right) B_0(M_W^2,M_W^2;p^2) 
+ {1\over 16 \pi^2} 4 M_W^2 
\Bigg\} \, ,  \label{def_SigmaT_A} \\
\Sigma_\A^L(p^2) & \equiv & 0 \label{def_SigmaL_A} \ . 
\eea

The tadpole contribution $A_0$ and the two-point integral $B_0$
which appear in Eqs.~(\ref{def_Sigma_H})--(\ref{def_SigmaL_A}) are
defined by the following equations: 
\bea 
A_0(M^2) & = & \int \ddq {1\over (q^2 + M^2)} \, , \\
B_0(M_1^2, M_2^2; p^2) & = & \int \ddq {1\over (q^2 + M_1^2)} {1\over
((q+p)^2 + M_2^2)} \ . 
\eea
Note that we are working in Euclidean space-time. 


\end{document}